\documentclass[preprintnumbers,article,amsmath,amssymb,floatfix,10pt,prd,onecolumn,
superscriptaddress,nofootinbib]{revtex4-2}
\usepackage{bm}
\usepackage{amsfonts}
\usepackage{latexsym}
\usepackage[latin1]{inputenc}
\usepackage{graphicx}
\usepackage{amsmath}
\usepackage{palatino}
\usepackage{mathpazo}
\usepackage{textcomp}
\usepackage{float}
\usepackage{booktabs}
\usepackage{dcolumn}
\usepackage{ragged2e}
\usepackage{hyperref}
\hypersetup{colorlinks,citecolor=blue}
\hypersetup{colorlinks=true,linkcolor=red,filecolor=magenta,    urlcolor=blue}
\usepackage{amsmath}
\usepackage{xcolor}
\usepackage{orcidlink}
\usepackage{epsfig}
\usepackage[caption=false]{subfig}
\usepackage{commath}

\usepackage{multirow}
\newcommand{\be}{\begin{equation}}
\newcommand{\ee}{\end{equation}}
\newcommand{\bea}{\begin{eqnarray}}
\newcommand{\eea}{\end{eqnarray}}
\newcommand{\eeas}{\end{eqnarray*}}
\newcommand{\beas}{\begin{eqnarray*}}

\def\jnl@style{\it}
\def\aaref@jnl#1{{\jnl@style#1}}

\def\aaref@jnl#1{{\jnl@style#1}}

\def\aj{\aaref@jnl{AJ}}                   
\def\apj{\aaref@jnl{ApJ}}                 
\def\apjl{\aaref@jnl{ApJ}}                
\def\apjs{\aaref@jnl{ApJS}}               
\def\apss{\aaref@jnl{Ap\&SS}}             
\def\aap{\aaref@jnl{A\&A}}                
\def\aapr{\aaref@jnl{A\&A~Rev.}}          
\def\aaps{\aaref@jnl{A\&AS}}              
\def\mnras{\aaref@jnl{Mon.~Not.~Roy.~Astron.~Soc.}}             
\def\prd{\aaref@jnl{Phys.~Rev.~D}}        
\def\prc{\aaref@jnl{Phys.~Rev.~C}}  
\def\prl{\aaref@jnl{Phys.~Rev.~Lett.}}    
\def\qjras{\aaref@jnl{QJRAS}}             
\def\skytel{\aaref@jnl{S\&T}}             
\def\ssr{\aaref@jnl{Space~Sci.~Rev.}}     
\def\zap{\aaref@jnl{ZAp}}                 
\def\nat{\aaref@jnl{Nature}}              
\def\aplett{\aaref@jnl{Astrophys.~Lett.}} 
\def\apspr{\aaref@jnl{Astrophys.~Space~Phys.~Res.}} 
\def\physrep{\aaref@jnl{Phys.~Rep.}}      
\def\physscr{\aaref@jnl{Phys.~Scr}}       
\def\commat{\aaref@jnl{Comm.~Math.~Phys.}}              
\def\science{\aaref@jnl{Science}}               
\def\cqg{\aaref@jnl{Classical Quant.~Grav.}}            
\def\jpcs{\aaref@jnl{JPCS}}                                     
\def\ijmpd{\aaref@jnl{Int.~J.~Mod.~Phys.~D}}                    
\def\grg{\aaref@jnl{Gen.~Relat.~Gravit.}}               
\def\rpp{\aaref@jnl{Rep.~Prog.~Phys.}}          
\def\npa{\aaref@jnl{Nucl.~Phys.~A}}        
\def\lrr{\aaref@jnl{Living Rev.~Rel.}}                   
\def\jcap{\aaref@jnl{J.~Cosmology Astropart.~Phys.}}    
\def\rmp{\aaref@jnl{Rev.~Mod.~Phys.}}   
\def\epjc{\aaref@jnl{Eur.~Phys.~J.~C}} 
\def\plb{\aaref@jnl{~Phy.~Lett.~B}} 
\def\mpla{\aaref@jnl{Mod.~Phy.~Lett.~A}} 
\def\arxiv{\aaref@jnl{arxiv.org}}

\allowdisplaybreaks[1]

\begin{document}

\color{black}       

\title{Periodic cosmic evolution in $f(Q)$ gravity formalism}

\author{Parbati Sahoo\orcidlink{0000-0002-5043-745X}}
\email{sahooparbati1990@gmail.com}
\affiliation{Astrophysics and Cosmology Research Unit, School of Mathematics, Statistics and Computer Science, University of KwaZulu--Natal, Private Bag X54001, Durban 4000, South Africa}
\affiliation{Department of Mathematics (SAS), Vellore Institute of Technology-Andhra Pradesh University, Andhra Pradesh - 522237, India}
\author{Avik De\orcidlink{0000-0001-6475-3085}}
\email{avikde@utar.edu.my}
\affiliation{Department of Mathematical and Actuarial Sciences, Universiti Tunku Abdul Rahman, Jalan Sungai Long, 43000 Cheras, Malaysia}
\author{Tee-How Loo\orcidlink{0000-0003-4099-9843}}
\email{looth@um.edu.my}
\affiliation{Institute of Mathematical Sciences, Faculty of Science,\\
Universiti Malaya, 50603 Kuala Lumpur, Malaysia}
\author{P.K. Sahoo\orcidlink{0000-0003-2130-8832}}
\email{pksahoo@hyderabad.bits-pilani.ac.in}
\affiliation{Department of Mathematics, Birla Institute of Technology and
Science-Pilani,\\ Hyderabad Campus, Hyderabad-500078, India.}

\begin{abstract}
We study the periodic cosmic transit behavior of accelerated universe in the framework of symmetric teleparallelism. The exact solution of field equations is obtained by employing a well known deceleration parameter (DP) called periodic varying deceleration parameter (PVDP), $q=m\cos k t-1$. The viability and physical reliability of the DP are studied by using the observational constraints. The dynamics of periodicity and singularity are addressed in details with respect to time and redshift parameter. Several energy conditions are discussed in this setting.  
\end{abstract}

\maketitle
\section{Introduction}\label{sec1}
The late time accelerated expansion is one of the prime research attentions today, and although observationally this phenomenon is well supported (Supernovae Ia, CMB, BAO and several other astrophysical measurements all confirm it), it still lacks a well-established theoretical root cause. The standard theory of gravity, introduced by Einstein in his general relativity (GR), fails to explain this without the assumption of yet undetected dark components. To overcome this situation, modification of GR took place and several new alternate gravity theories drew the research focus while trying to describe the accelerated expansion as a geometric effect. Other than the curvature, there are other two fundamental geometric entities attributed to gravity, the torsion and non-metricity of flat spacetime. On a flat space, one can construct the so-called torsion scalar $\mathbb{T}$ from the torsion tensor in the metric teleparallel theory and the non-metricity scalar $Q$ in the symmetric teleparallel theory. Thereafter, much like the Lagrangian $\mathcal{L}=\sqrt{-g}R$ ($R$ being the Ricci scalar) in the Einstein-Hilbert action $S=\int{d^4x}\mathcal{L}$ in GR, we swap $R$ by $\mathbb{T}$ in the metric teleparallel theory and by $Q$ in the symmetric teleparallel theory to obtain the respective field equations. However, the latter two theories are equivalent to GR up to a boundary term and also facing the same dependencies on dark components as GR. Therefore, extended $f(\mathbb{T})$ and $f(Q)$-theories were formulated in the metric teleparallel and symmetric teleparallel, the same way as the modified $f(R)$-theory was developed from the Einstein-Hilbert action in the realm of GR. In the present work, we concentrate on the newer $f(Q)$ theory of gravity which recently is centre of much research attention, see \cite{fQ} and the references therein for a detailed list of publications in this theory.     

We model the present universe by spatially isotropic and homogeneous Friedman-Lema\^{i}tre-Robertson-Walker (FLRW) spacetime. This spacetime, mathematically, is a warped product of time and a 3-dimensional spaceform, and the warping function $a(t)$, popularly known as the scale factor of the universe, controls the expansion of the universe. There are several important cosmological parameters formulated from the scale factor which help us in understanding the dynamics of the universe. The main two of them are the Hubble parameter $H(t)=\frac{\dot{a}}{a}$, demonstrating the expansion or contraction of the universe and the deceleration parameter (DP) $q=-\frac{a\ddot{a}}{\dot{a}^2}$, demonstrating the acceleration or deceleration of the universe. For an accelerating universe we have $\ddot{a}>0$ reflecting in a negative DP. The negative sign in the definition of the DP is significant in the sense that it was originally defined for a decelerating ($\ddot{a}<0$) universe, pre-dark energy dominated accelerating era of the universe and thus a negative sign was required at that time in the definition to make $q$ ultimately positive. But after the well-documented observational evidence of the accelerating universe, the present day value of the DP $q_0<0$. The DP is varying with respect to cosmological time, in general, unless in some particular models we restrict it to be constant \cite{ber}. In \cite{ber}, the authors actually considered some observationally consistent law of variation of Hubble parameter which produced such a constant $q$, and this also possibly could approximate slowly time-varying cases. In the standard $\Lambda$CDM model, $q$ is varying between $\frac{1}{2}$ and $-1$ and thus the DP as a linear function of temporal coordinate $t$ or redshift $z$ was considered next \cite{linear}, followed by a quadratic representation \cite{bakry} and also possibly as a generalized Taylor series expansion. The variation law we consider for the DP here is of periodic type,  which originally was analyzed in \cite{pvdp} for GR and later in the modified $f(R,T)$-theory \cite{sahoo/pvdp}.

In the present manuscript, we investigate the dynamics of FLRW universe, scrutinising it from observational viewpoint and try to find the answer whether a PVDP in modified $f(Q)$-theory supports the present dataset. After introduction, we give a brief description of the formulation of $f(Q)$-theory in section \ref{sec2}, and that of PVDP $q$ in section \ref{sec3}. The dynamics of this model and ECs are considered in the next two sections, \ref{sec4} and \ref{sec5} respectively, followed by the kinematical properties in section \ref{sec6}. The whole study is concluded in the last section \ref{sec7}.  

\section{Formulation in $f(Q)$-theory}\label{sec2}
In this section, we discuss about the basic formulation of the $f(Q)$ theory. If we consider our base connection in a smooth manifold $(M,g)$ to possess both metric-compatibility and torsion-free conditions, then there is only a unique connection available, the Levi-Civita connection $\mathring{\Gamma}$ and it complies with the metric $g$ by
\begin{equation}
\mathring{\Gamma}^\alpha_{\,\,\,\mu\nu}=\frac{1}{2}g^{\alpha\beta}\left(\partial_\nu g_{\beta\mu}+\partial_\mu g_{\beta\nu}-\partial_\beta g_{\mu\nu}  \right).
\end{equation}
So, the Levi-Civita connection is not an independent contributor in the spacetime geometry and merely a function of the metric $g$. The situation changes once we consider an affine connection $\Gamma$ on a flat spacetime which is not metric-compatible, that is, the covariant derivative $\nabla$ of the metric $g$ is non-zero. This produces the non-metricity tensor 
\begin{equation} \label{Q tensor}
Q_{\lambda\mu\nu} := \nabla_\lambda g_{\mu\nu}=\partial_\lambda g_{\mu\nu}-\Gamma^{\beta}_{\,\,\,\lambda\mu}g_{\beta\nu}-\Gamma^{\beta}_{\,\,\,\lambda\nu}g_{\beta\mu}\neq 0 \,,
\end{equation}
We can write
\begin{equation} \label{connc}
\Gamma^\lambda{}_{\mu\nu} := \mathring{\Gamma}^\lambda{}_{\mu\nu}+L^\lambda{}_{\mu\nu}
\end{equation}
where $L^\lambda{}_{\mu\nu}$ is the disformation tensor. Then we can prove that
\begin{equation} \label{L}
L^\lambda{}_{\mu\nu} = \frac{1}{2} (Q^\lambda{}_{\mu\nu} - Q_\mu{}^\lambda{}_\nu - Q_\nu{}^\lambda{}_\mu) \,.
\end{equation}
We can construct two different types of non-metricity vectors,
\begin{equation*}
 Q_\mu := g^{\nu\lambda}Q_{\mu\nu\lambda} = Q_\mu{}^\nu{}_\nu \,, \qquad \tilde{Q}_\mu := g^{\nu\lambda}Q_{\nu\mu\lambda} = Q_{\nu\mu}{}^\nu \,.
\end{equation*}
The non-metricity conjugate tensor $P^\lambda{}_{\mu\nu}$ is given by
\begin{equation} \label{P}
P^\lambda{}_{\mu\nu} = \frac{1}{4} \left( -2 L^\lambda{}_{\mu\nu} + Q^\lambda g_{\mu\nu} - \tilde{Q}^\lambda g_{\mu\nu} -\frac{1}{2} \delta^\lambda_\mu Q_{\nu} - \frac{1}{2} \delta^\lambda_\nu Q_{\mu} \right) \,.
\end{equation} 

The central character of this theory, the non-metricity scalar $Q$ is defined as
\begin{equation} \label{Q}
Q=-Q_{\alpha\beta\gamma}P^{\alpha\beta\gamma}.
\end{equation}
As discussed in \cite{gde}, the $f(Q)$-theory was constructed by using the constraints, $R^\rho{}_{\sigma\mu\nu} = 0$. That means there exist a special coordinate system such that the affine connection vanishes, $\Gamma^\lambda{}_{\mu\nu} = 0$. This situation is called the coincident gauge. Under this circumstance, the metric is the only dynamical variable. But as mentioned in \cite{zhao}, in any other coordinate such that the connection does not vanish, the evolution of the metric will be affected, and results into a completely different theory. Therefore, by varying the action term
\begin{equation*}
S = \int \left[-\frac{1}{2}f(Q) + \mathcal{L}_M \right] \sqrt{-g}\,d^4 x
\end{equation*}
with respect to the metric, we obtain the field equations 
\begin{equation} \label{FE1}
\frac{2}{\sqrt{-g}} \nabla_\lambda (\sqrt{-g}f_QP^\lambda{}_{\mu\nu}) +\frac{1}{2}f g_{\mu\nu} + f_Q(P_{\nu\rho\sigma} Q_\mu{}^{\rho\sigma} -2P_{\rho\sigma\mu}Q^{\rho\sigma}{}_\nu) = T_{\mu\nu}
\end{equation}
where $f_Q = \partial f/\partial Q$ and $T_{\mu\nu}$ is the energy-momentum tensor generated from the matter Lagrangian $\mathcal{L}_M$. However, this equation is only valid in the coincident gauge coordinate. In the current discussion, we consider the FLRW universe, with line element given in terms of Cartesian coordinates which is automatically a coincident gauge
\begin{equation}\label{RW}
ds^2=-dt^2+a^2(t)(dx^2+dy^2+dz^2).
\end{equation}
Taking into account the perfect fluid energy-momentum tensor
$T_{\mu\nu}=(\rho+p)u_\mu u_\nu+pg_{\mu \nu}$, where $\rho$ and $p$ are energy density and isotropic pressure with equation of state (EOS) parameter $\omega$ satisfying $p=\omega \rho$, the equations of motion are obtained as
\begin{equation}\label{rho1}
\rho=-\frac{f}{2}+6H^2f_Q  ,
\end{equation}
and 
\begin{equation}\label{p1}
p=\frac{f}{2}-6H^2f_Q -24H^2\dot{H}f_{QQ}-2 \dot{H}f_Q,
\end{equation}
where the over dot stands for differentiation with respect to $t$. To find an exact solution to this undetermined system of equations, we make an adhoc assumption with one of the unknowns $H$, or more explicitly, in terms of $q\left (=-1-\frac{\dot{H}}{H^2}\right)$ in the following section.

\section{Periodically Varying Deceleration Parameter (PVDP)\label{sec3}}
\begin{figure}[H]
\centering
\begin{minipage}{82mm}
\includegraphics[width=85 mm]{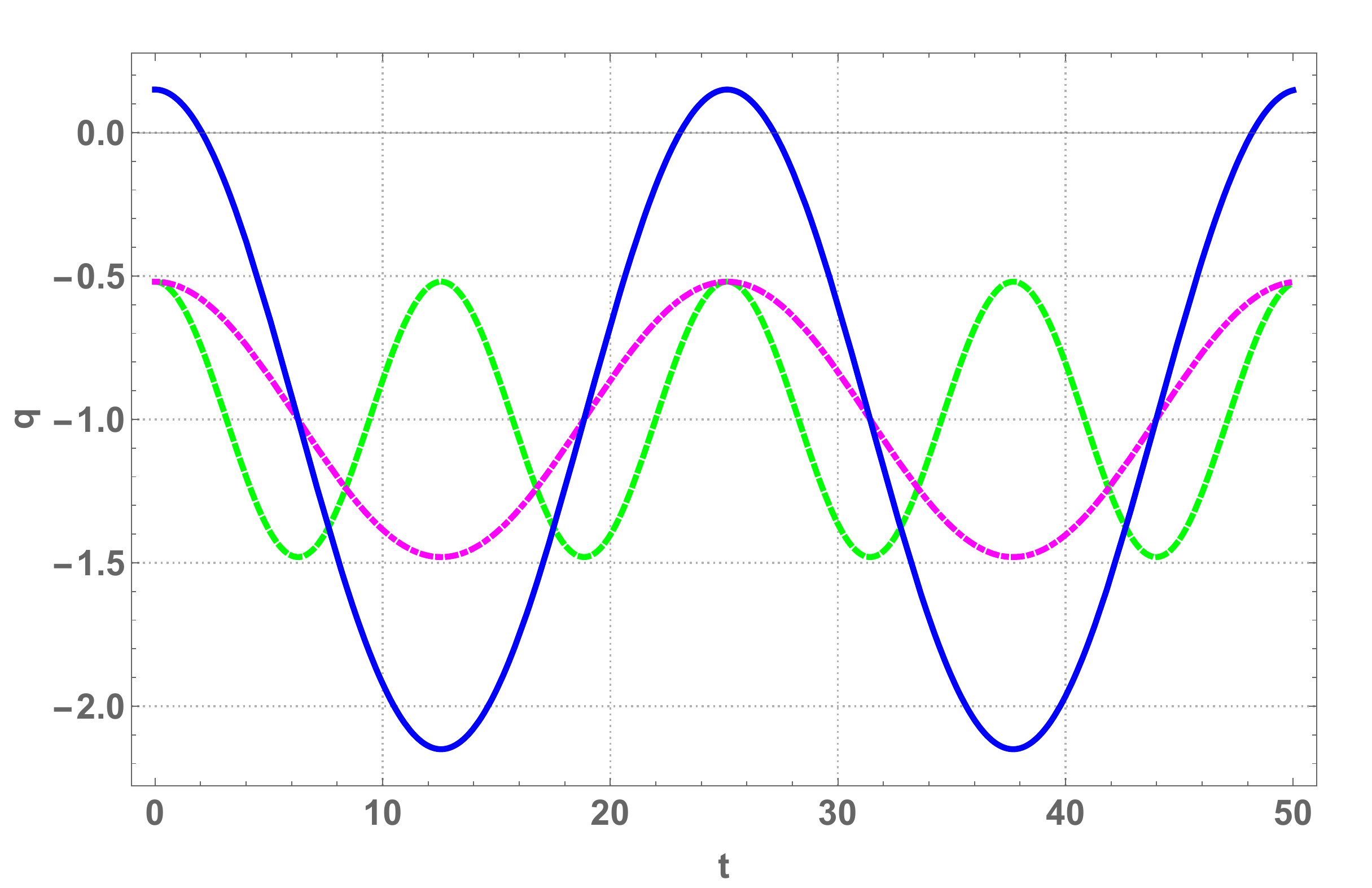}

\end{minipage}
\hfill
\begin{minipage}{82mm}
\includegraphics[width=85 mm]{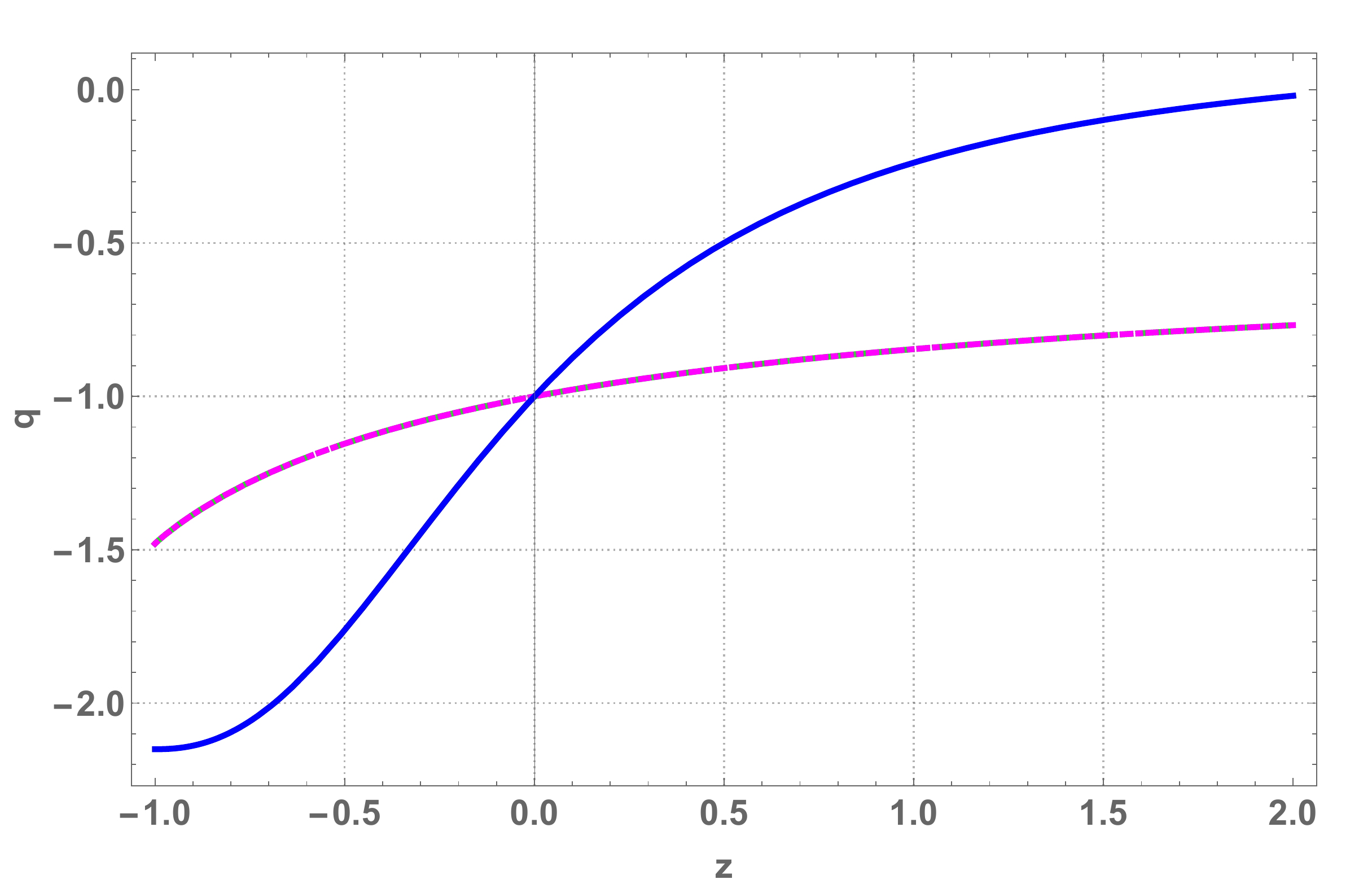}
  \end{minipage}
  \caption{Evolution of PVDP as a function of time (in $Gyr$) [left] and redshift [right] for  ($m$, $k$)=($0.480012$, $0.5$)  [green curve], and ($m$, $k$)= ($0.480003$, $0.25$) [magenta curve]. The cosmic transit behaviour is obtained for $m>1$ with ($m$, $k$)=($1.15$, $0.1$)  [blue curve].}\label{PVDP}
\end{figure}

Deceleration parameter $q$ takes a vital role in describing the dynamics of the universe, whether it is late time accelerated phase or past decelerated phase of the universe; both cosmic transit behaviors can be speculated from signature flipping nature of the DP (see \cite{sahoo/pvdp} and references therein). We consider the following special form of $q$ given by \cite{pvdp,sahoo/pvdp}
\begin{equation}\label{deceleration}
q=m\cos kt-1,
\end{equation}
where the positive constant $m$ determine the peaks of PVDP and positive $k$ decides the periodicity of the PVDP. These evolutionary phenomena can be observed through graphical representation given in figure (\ref{PVDP}), where the change of DP with redshift $z<2$ is plotted. The universe  decelerates with a positive DP $q=m-1$  (for $m>1$) at an initial epoch, and evolves into a negative peak of $q=m-1$ (for $m<1$) and attends super-exponential expansion with $q=-m-1$. After that, it again increases and comes back to the initial state. The evolutionary behaviour of $q$ is periodically repeated. As per the current scenario, the present value of DP  and Hubble parameter are $q_0=-0.52$ and $H_0=69.2$ \cite{Ratra/18, Amirhashchi/20}. With these constraints the equation (\ref{deceleration}) yields a relation between $k$ and $m$ as $k=H_0 \cos^{-1} (\frac{q_0+1}{m})$. By using this relationship we  obtain the value of $m$ for different values of $k$. We consider two sets of values for $m$ and $k$ from the above relationship i.e. ($m$, $k$)=($0.480012$, $0.5$), ($0.480003$, $0.25$)) and the 3rd choice was arbitrarily taken to experience the complete transition from decelerating to accelerating phase. However, the first two choices are leading to the accelerating phase in late time era. The complete evolution is depicted in the following plots. 

Therefore, in present model the universe starts with a decelerating phase and evolves into a phase of super-exponential expansion in a cyclic history. The cyclic transitional behavior is addressed for $m \geq 1$. It maintains the accelerating phase at $0<m<1$ and leads the present value of $q$ at $m\simeq 0.48$.  
The corresponding Hubble parameter obtained as
\begin{equation}\label{Hubble}
H=\frac{k}{m\sin kt +\lambda},
\end{equation}
where $\lambda$ is an integrating constant. One can observe here that the singularity appears when denominator vanishes for $|\lambda| \leq m$. Similarly, the universe is contracting for $\lambda <-m$ and to avoid such kind of issues, it is worth to consider $\lambda > m$. However, without loss of generality and for simplicity we have considered $\lambda=0$ and the Hubble function becomes 
\begin{equation}
H=\frac{k}{m\sin kt}.\label{eq:9}
\end{equation}

The cosmic scale factor  $a$ under the ansatz of PVDP is obtained by integrating the Hubble function in equation \eqref{eq:9} as
\begin{equation}
a=a_0\left[\tan \left(\frac{1}{2}kt\right)\right]^{\frac{1}{m}},\label{eq:10}
\end{equation}
where $a_0$ is the scale factor at the present epoch  and can be taken as 1. Inverting equation \eqref{eq:10}, we obtain
\begin{equation}
t=\frac{2 \tan ^{-1}\left[\frac{1}{(z+1)^m}\right]}{k}.\label{eq:11}
\end{equation}
 We use the above choices of $m$ and $k$ hereafter to describe the dynamical features of the model through numerical plots.
\begin{figure}[H]
\centering
\begin{minipage}{82mm}
\includegraphics[width=85 mm]{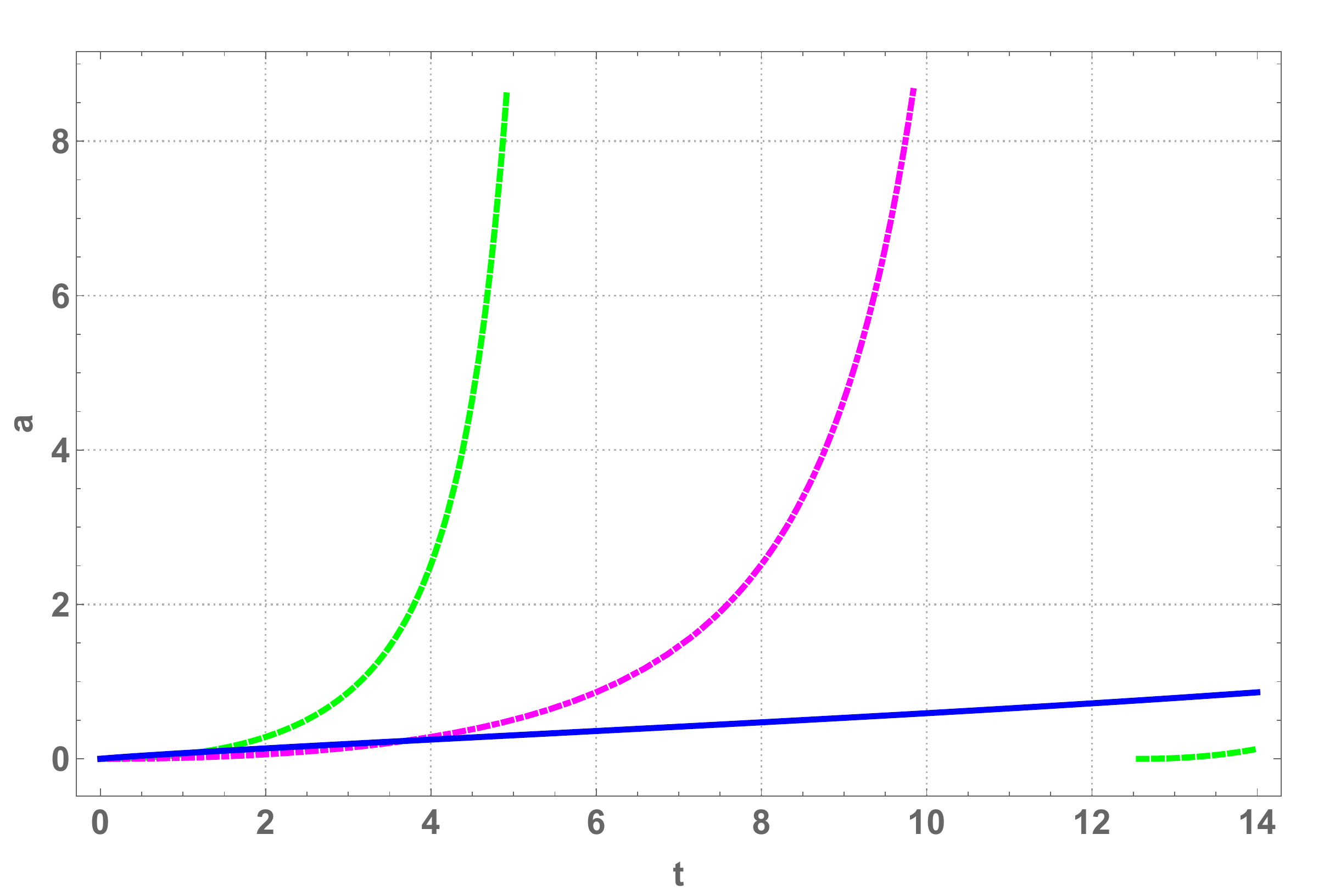}
  \caption{Scale factor as a function of time (in $Gyr$) for  ($m$, $k$)=($0.480012$, $0.5$)  [green curve], ($m$, $k$)= ($0.480003$, $0.25$) [magenta curve], and ($m$, $k$)=($1.15$, $0.1$)  [blue curve].}\label{fig3}
\end{minipage}
\hfill
\begin{minipage}{82mm}
\includegraphics[width=85 mm]{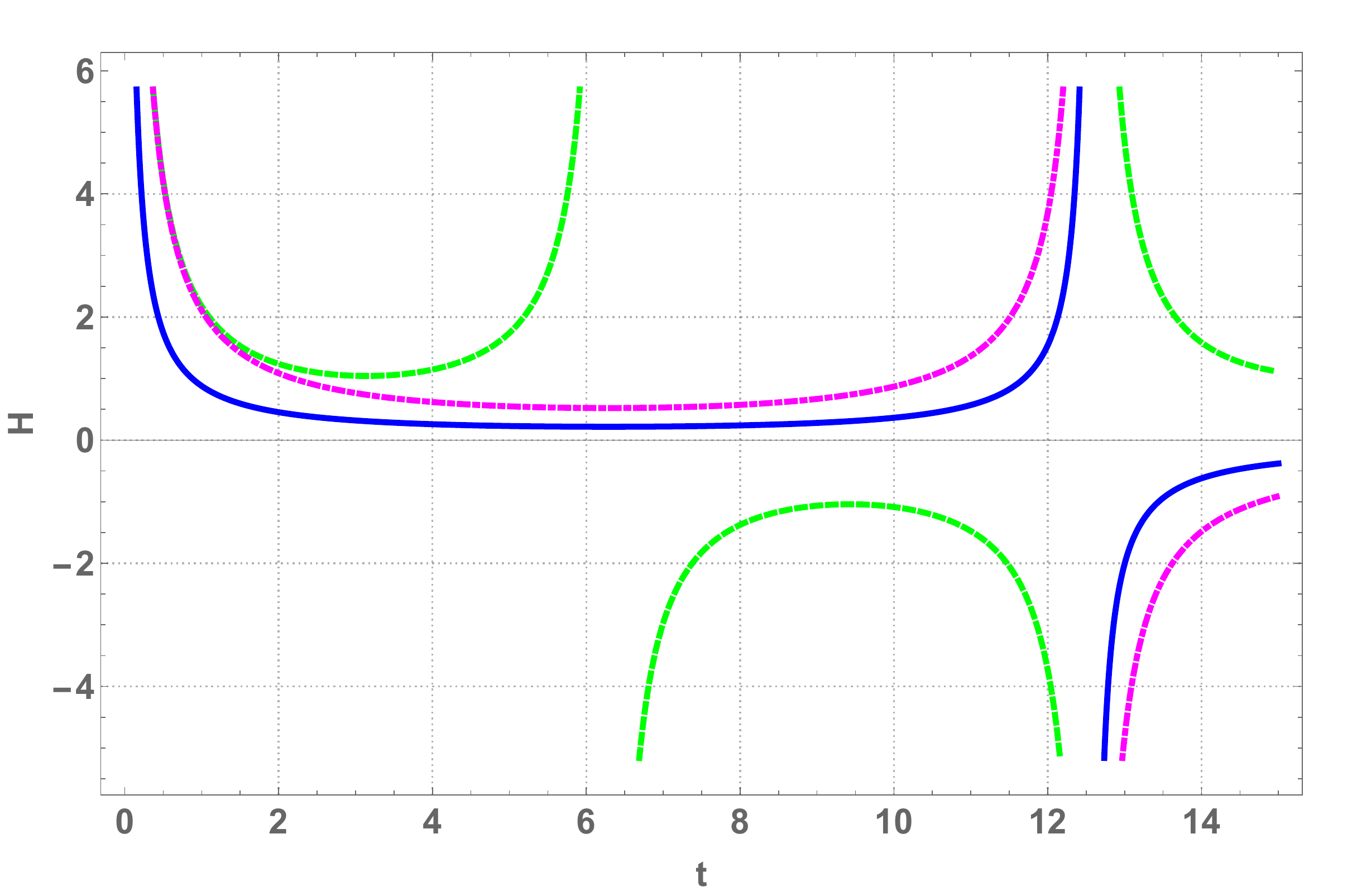}
  \caption{Hubble parameter as a function of time (in $Gyr$) for  ($m$, $k$)=($0.480012$, $0.5$)  [green curve], ($m$, $k$)= ($0.480003$, $0.25$) [magenta curve], and ($m$, $k$)=($1.15$, $0.1$)  [blue curve].}\label{fig4}
  \end{minipage}
\end{figure}

\section{Dynamical Properties of the Model}\label{sec4}
Our physical understanding of the universe can be extended through the recent scenario of cosmic expansion and nature of the mechanism behind this acceleration. In this work, we assume parametrization of DP and study the physical reliability of the model in the framework of $f(Q)$ gravity. In order to check the model's viability, the dynamical properties are proposed here.  

From (\ref{rho1}) and (\ref{p1}) thus we obtain the density ($\rho$), pressure ($p$), and EoS parameter ($\omega=\frac{p}{\rho}$) are
\begin{eqnarray}
\rho&=&-\frac{f}{2}+6 f_Q \left(\frac{k}{m \sin (k t)}\right)^2,\\
 p&=&\frac{f}{2}+\frac{4 k^2 \csc (k t) \left[\cot (k t) \left(f_Q m^2+12 f_{QQ} k^2 \csc ^2(k t)\right)-3 f_Q m \csc (k t)\right]}{2 m^3},\\
\omega&=& \frac{4 k^2 \cot (k t) \csc (k t) \left(f_Q m^2+12 f_{QQ} k^2 \csc ^2(k t)\right)}{-f m^3+12 f_Q k^2 m \csc ^2(k t)}-1.
\end{eqnarray}

With the choice of $f(Q)=\alpha Q^{n+1}+\beta$, the above explicit expressions read
\begin{equation}\label{rho-1}
\rho=\frac{\alpha 6^{n+1} (2 n+1)}{2} \left(\frac{k^2 \csc ^2(k t)}{m^2}\right)^{n+1}-\frac{\beta}{2},
\end{equation}
\begin{equation}\label{p-1}
p=\frac{\beta }{2 }+\alpha  6^{n}  (2 n+1) (2 m (n+1) \cos (k t)-3) \left(\frac{k^2 \csc ^2(k t)}{m^2}\right)^{n+1},
\end{equation}

\begin{equation}\label{omega-1}
\omega=-\frac{\alpha  m^2 2^{n+1} 3^n (2 n+1) (2 m (n+1) \cos (k t)-3) \left(\frac{k^2 \csc ^2(k t)}{m^2}\right)^{n+1}+\beta  m^2}{\beta  m^2-\alpha  m^2 6^{n+1} (2 n+1) \left(\frac{k^2 \csc ^2(k t)}{m^2}\right)^{n+1}},
\end{equation}
where $\alpha$, $\beta$, and $n$ are constants.\\

To obtain the qualitative behaviour of the physical parameters in $f(Q)$ gravity 
in the next subsection, we shall analyze the model parameters $\alpha,\ \  \beta \ \ $ and $\ \ n$. For the choice of parameters $\alpha=-1 \ \ $ and  $\ \ \beta=0=n$ our proposed model reduces to GR. Here, we consider the simplest linear form of f(Q) function $f(Q)=\alpha Q+\beta$ with $n=0$ and compared the results with the non-linear form $f(Q)=\alpha Q^2+\beta$ with $n=1$.
\\

\begin{figure}[H]
\centering
\begin{minipage}{55mm}
\includegraphics[width=60 mm]{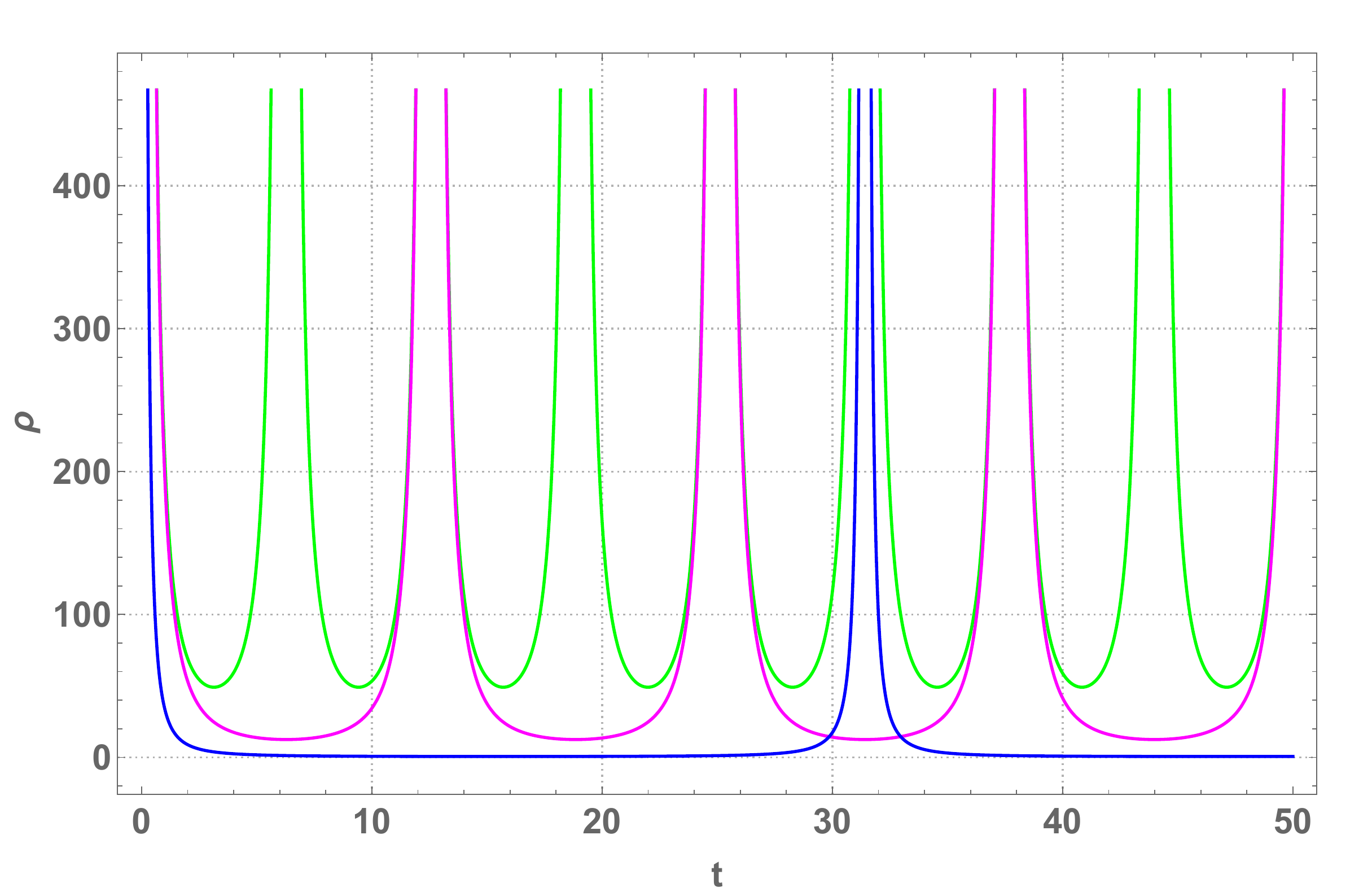}
\end{minipage}
\hfill
\begin{minipage}{55mm}
\includegraphics[width=60 mm]{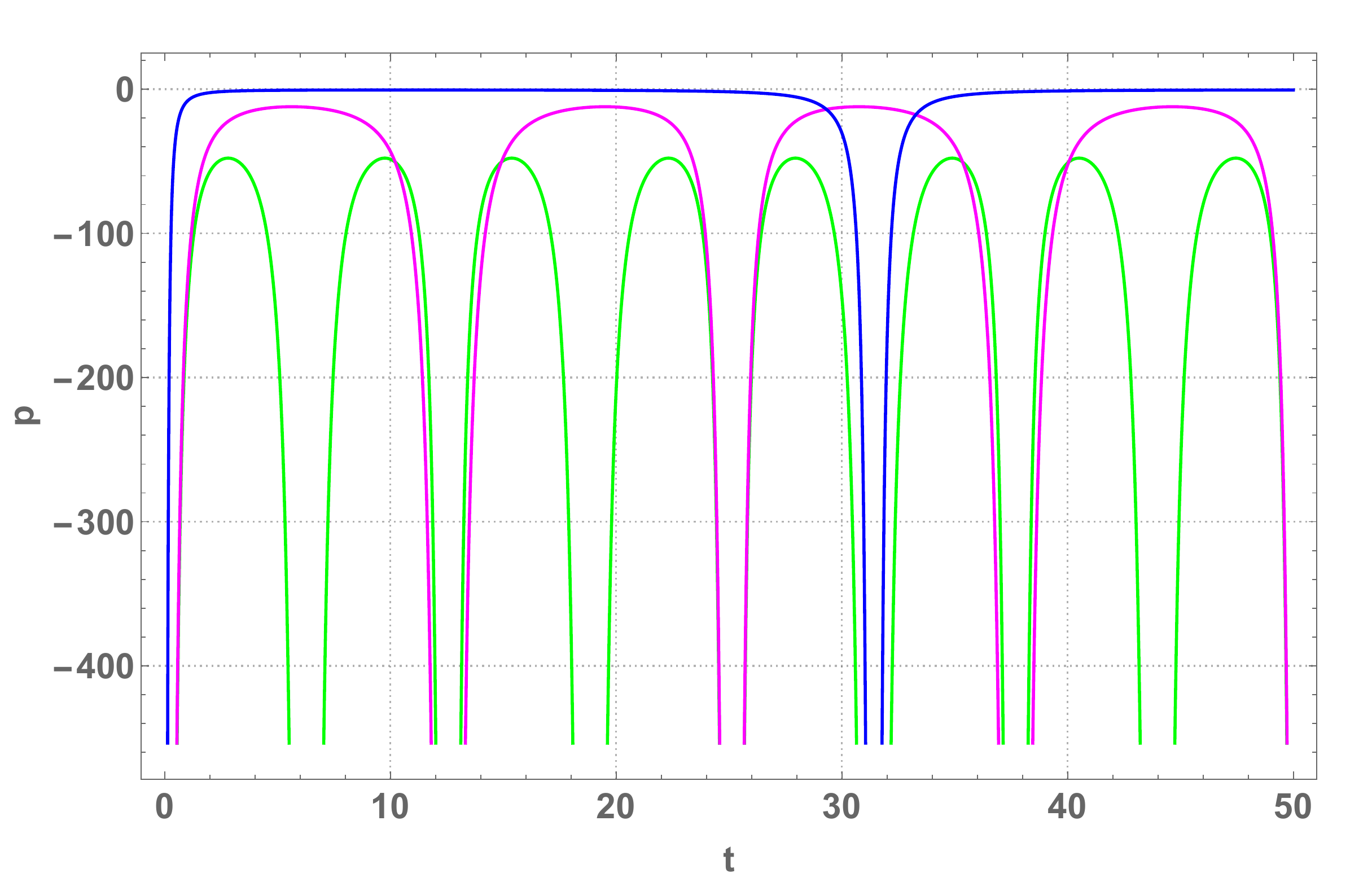}
\end{minipage}
\hfill
\begin{minipage}{55mm}
\includegraphics[width=60 mm]{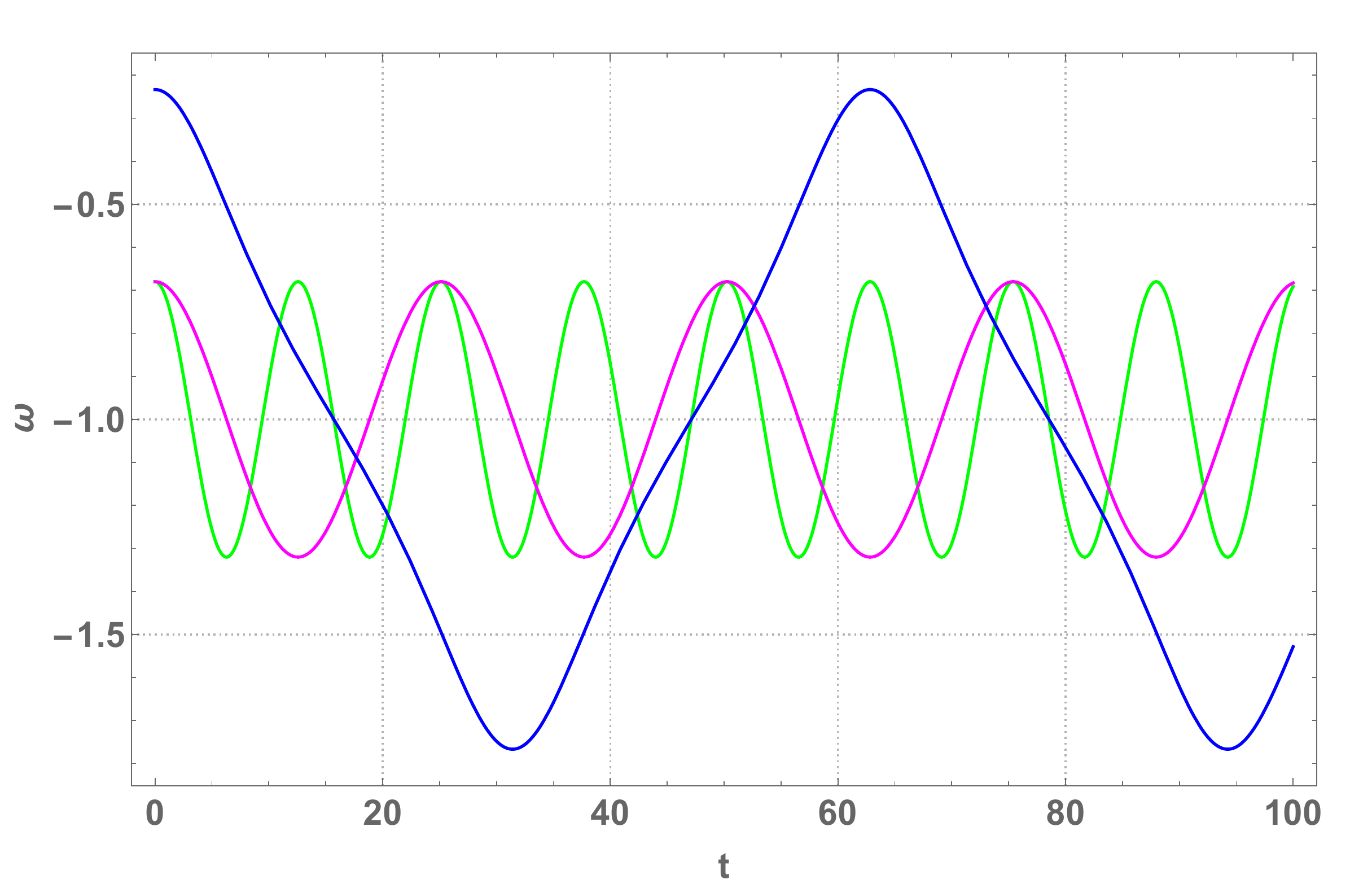}
\end{minipage}
\caption{Time (in $Gyr$) evolution of $\rho$, $p$, and $\omega$ with $\alpha=15$, $\beta=-0.5$, $n=0$  for  ($m$, $k$)=($0.480012$, $0.5$)  [green curve], ($m$, $k$)= ($0.480003$, $0.25$) [magenta curve], and ($m$, $k$)=($1.15$, $0.1$)  [blue curve]. }\label{figcase1t}
\end{figure}

\begin{figure}[H]
\centering

\begin{minipage}{55mm}
\includegraphics[width=60 mm]{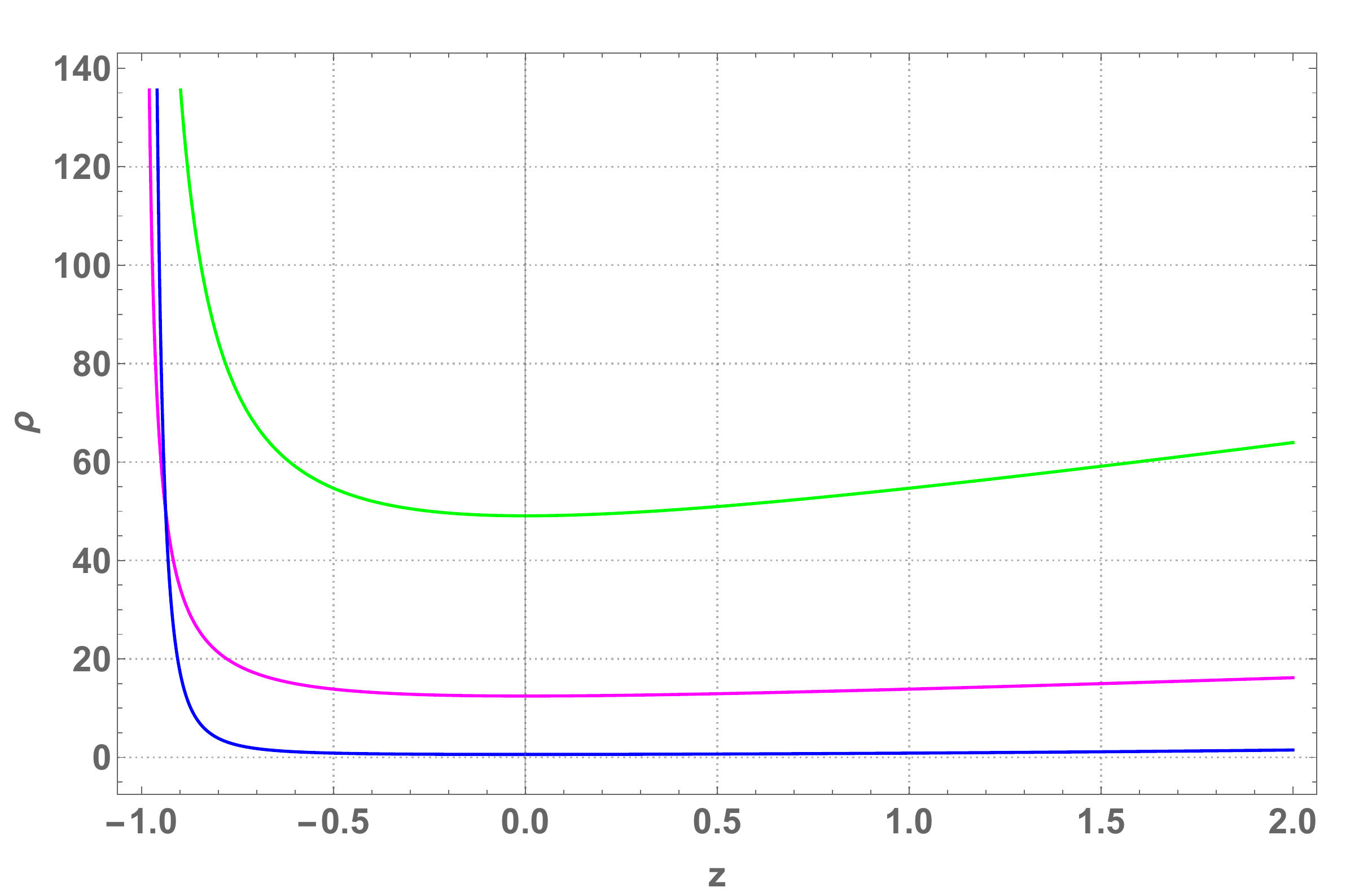}
\end{minipage}
\hfill
\begin{minipage}{55mm}
\includegraphics[width=60 mm]{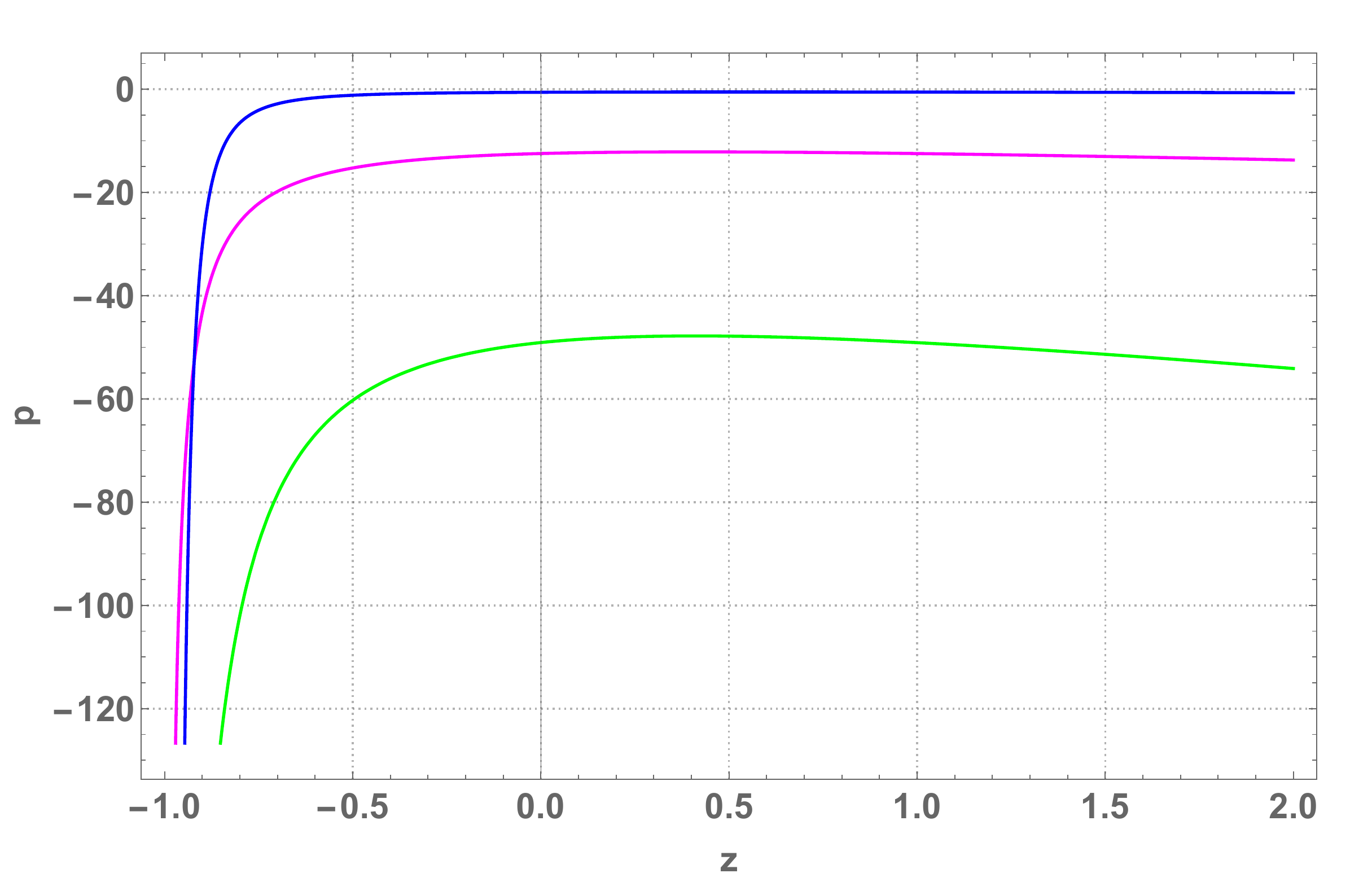}
\end{minipage}
\hfill
\begin{minipage}{55mm}
\includegraphics[width=60 mm]{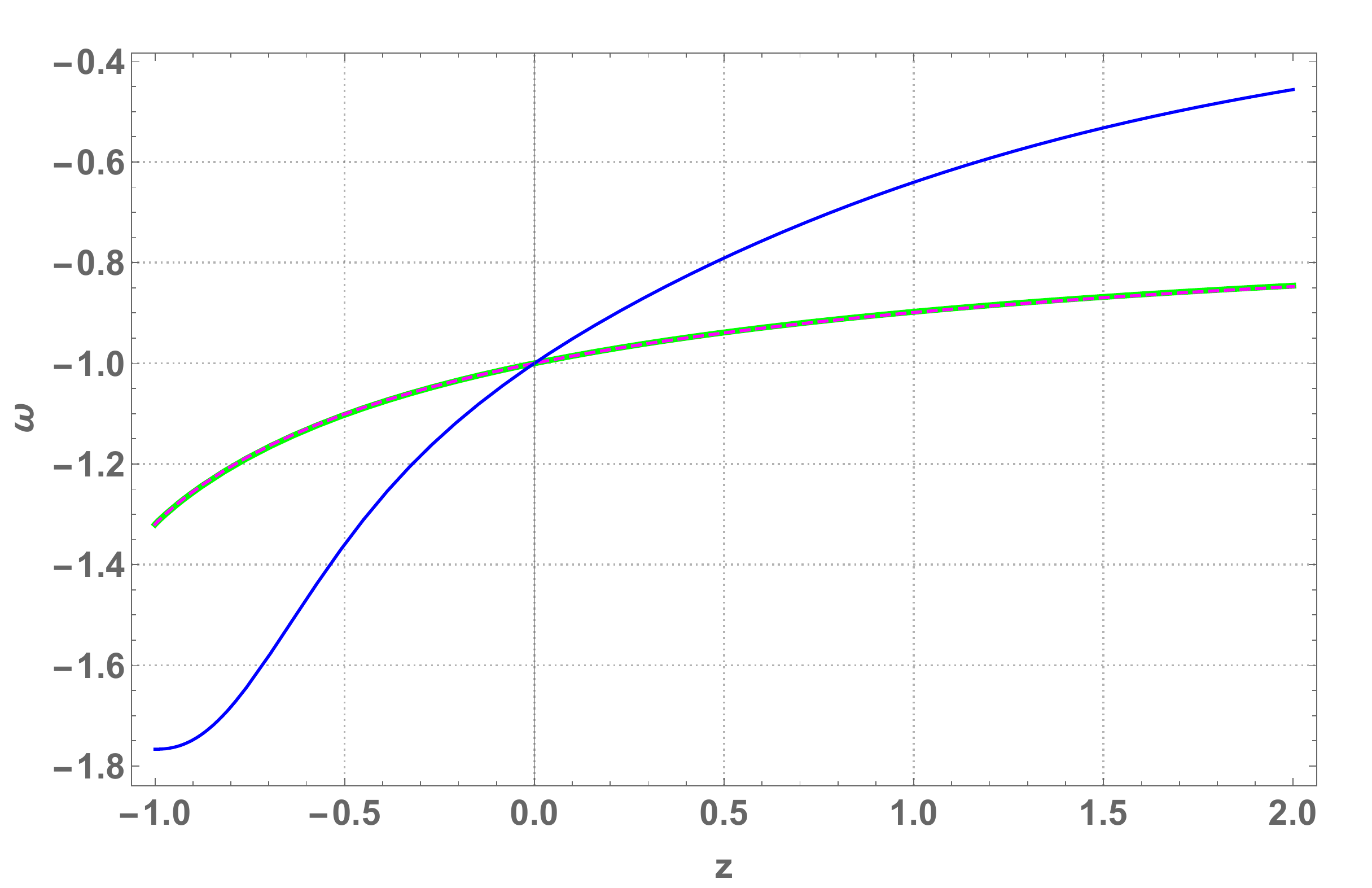}
\end{minipage}
\caption{Redshift evolution of $\rho$, $p$, and $\omega$ with $\alpha=15$, $\beta=-0.5$, $n=0$ for  ($m$, $k$)=($0.480012$, $0.5$)  [green curve], ($m$, $k$)= ($0.480003$, $0.25$) [magenta curve], and ($m$, $k$)=($1.15$, $0.1$)  [blue curve]. } \label{figcase1z}
\end{figure}

\textit{\textbf{Case--1: ($n=0$) }} This model is in linear form  for $n=0$ i.e. $f(Q)=\alpha Q+\beta$. After inserting it in the equation (\ref{rho-1}-\ref{omega-1}) the physical behavior with respect to time component and redshift parameter are depicted in the figures (\ref{figcase1t}--\ref{figcase1z}) for $\alpha=15$ and $\beta=-0.5$.

In this choice of $f(Q)$ gravity, we are experiencing the big rip singularity at finite time $t=\frac{j\pi}{k}$, where $j$ is an integer (j=0,1,2,3.....).  The periodic time evolution of pressure and energy density with finite time singularity are portrayed in figure profile  (\ref{figcase1t}). The pressure profile lies in accelerating phase with negative values for each value $m$ and $k$. On the other hand, the positivity of energy density depends on positive value of $\alpha$. For the different choices of $k$, the cosmic singularity occurs in different time period corresponding to $k$. For example; the time period, $t=0, 31.4, 62.8, \cdots$ for the value of $k=0.1$. The interesting feature is that, in a given cosmic cycle, it starts from a very large value at an initial time ($t\rightarrow 0$) and decreases to a minimum, $\rho_{min}$, and then again increases with the growth of time. The minimum in energy density occurs at a time given by $t=\frac{(n+1)\pi}{2k}$.
 However, the EoS parameter is transitioning periodically without any finite time singularity within the range between $-2.5< \omega<0.5$ for the choice of $(m, k)=(1.15, 0.1)$ and $-1.5< \omega<-0.5$ for the choice of $(m, k)=((0.480012, 0.5), (0.480003, 0.25)) $ . The similar behavior with respect to redshift parameter can be observed in the figure profile (\ref{figcase1z}). \\

\begin{figure}[H]
\centering
\begin{minipage}{55mm}
\includegraphics[width=60 mm]{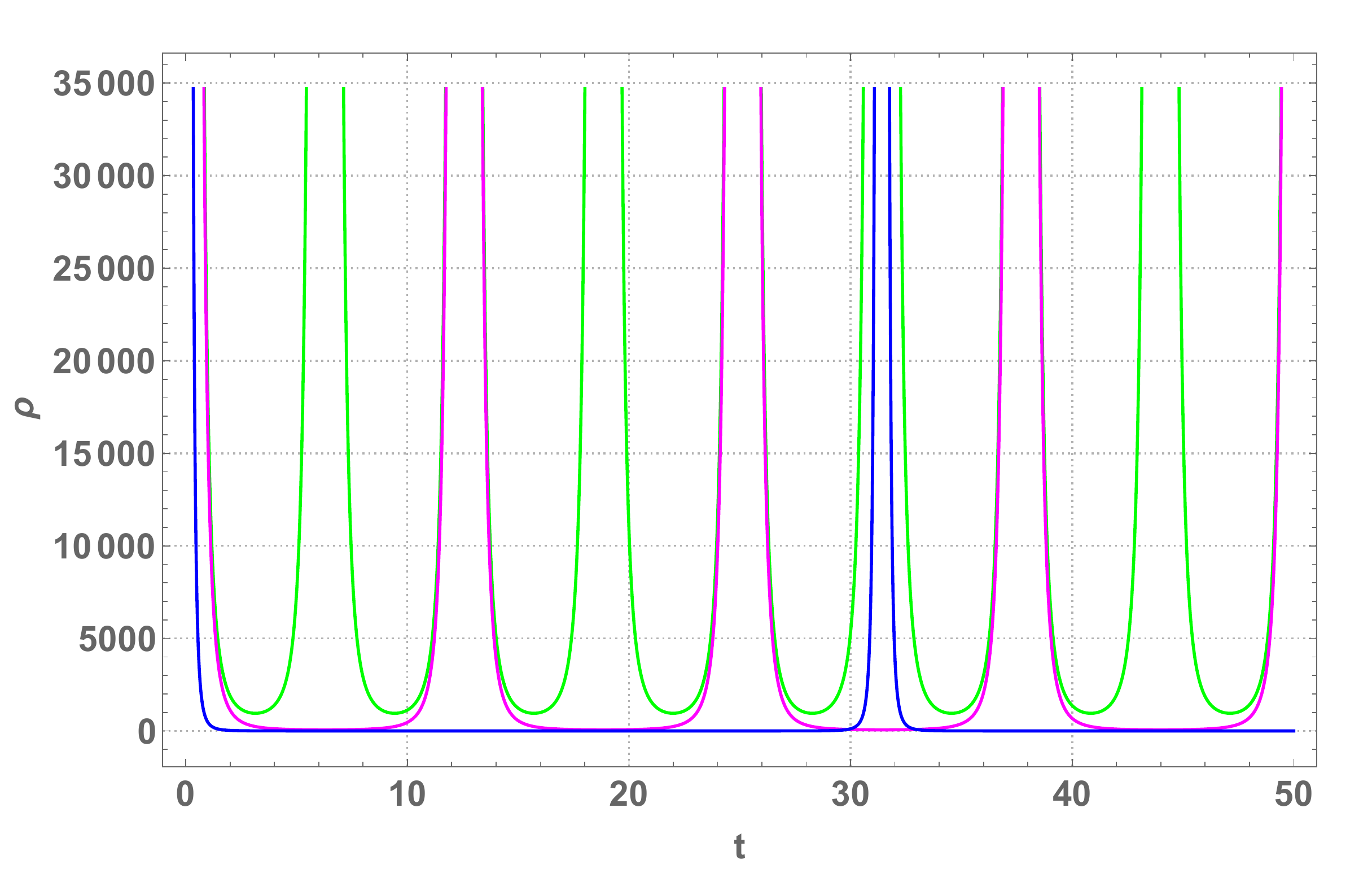}
\end{minipage}
\hfill
\begin{minipage}{55mm}
\includegraphics[width=60 mm]{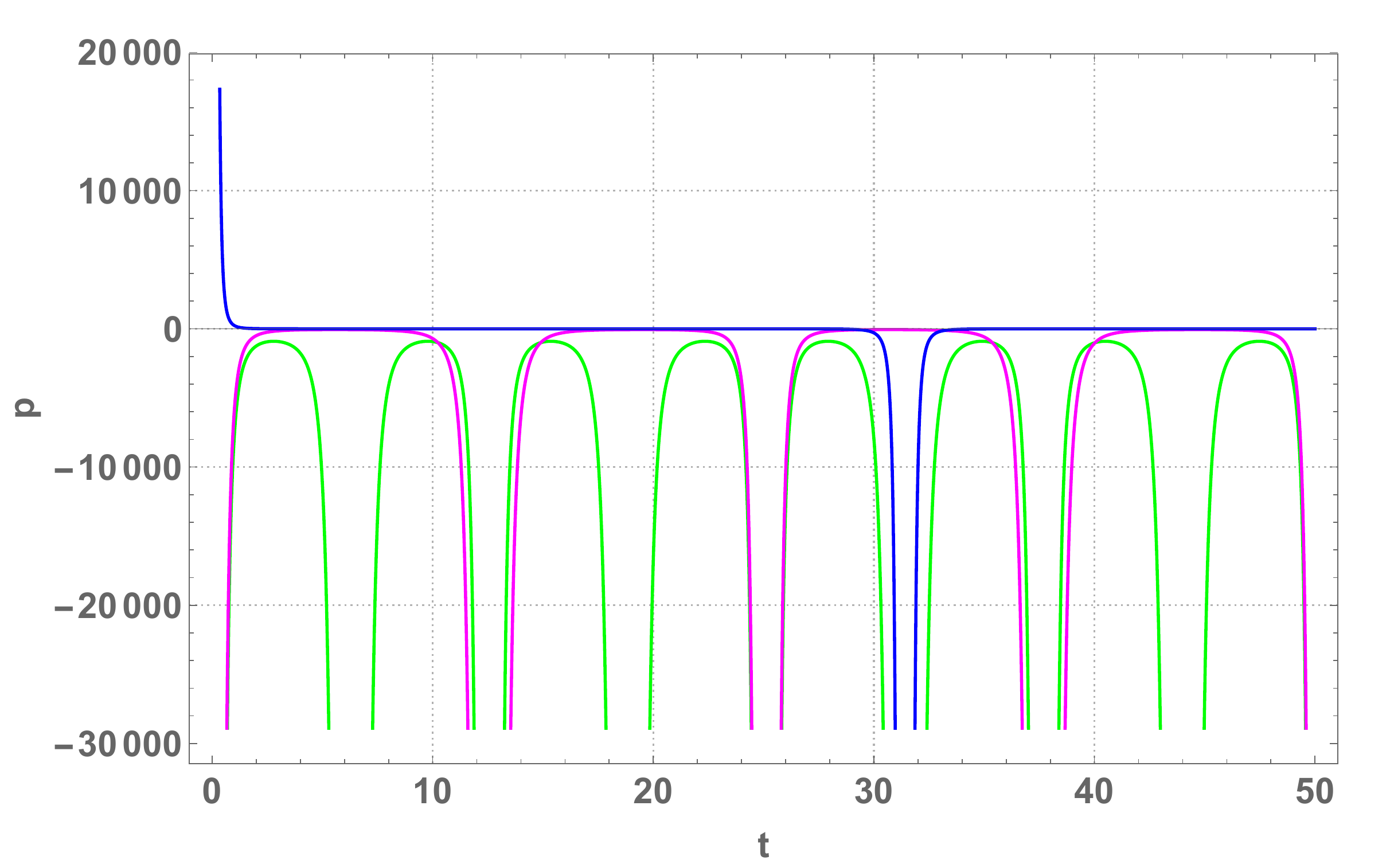}
\end{minipage}
\hfill
\begin{minipage}{55mm}
\includegraphics[width=60 mm]{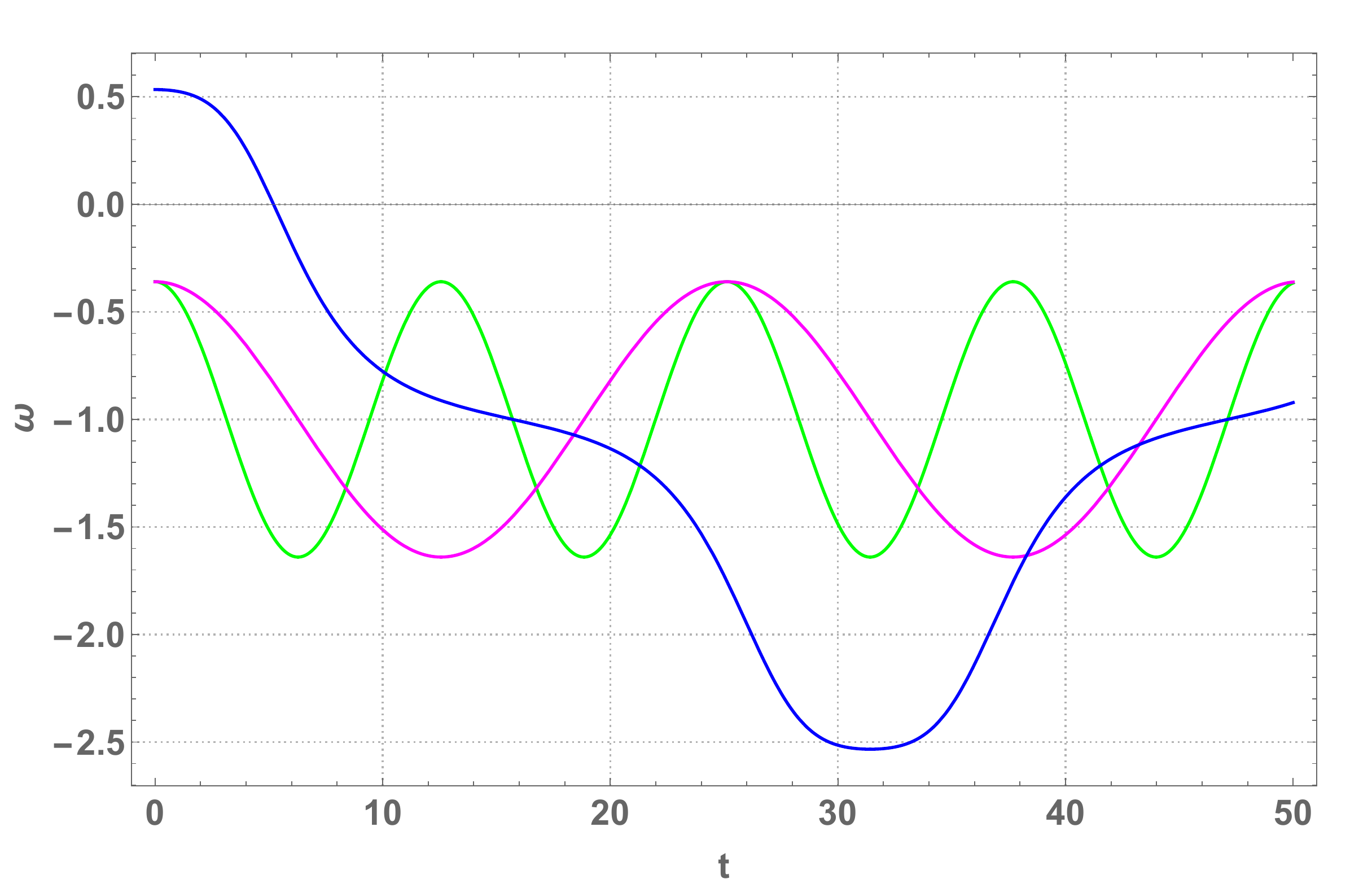}
\end{minipage}
\caption{Time (in $Gyr$) evolution of $\rho$, $p$, and $\omega$ with $\alpha= 15$, $\beta= -0.5$, $n=1$  for  ($m$, $k$)=($0.480012$, $0.5$)  [green curve], ($m$, $k$)= ($0.480003$, $0.25$) [magenta curve], and ($m$, $k$)=($1.15$, $0.1$)  [blue curve]. }\label{figcase2t}
\end{figure}

\begin{figure}[H]
\centering
\begin{minipage}{55mm}
\includegraphics[width=60 mm]{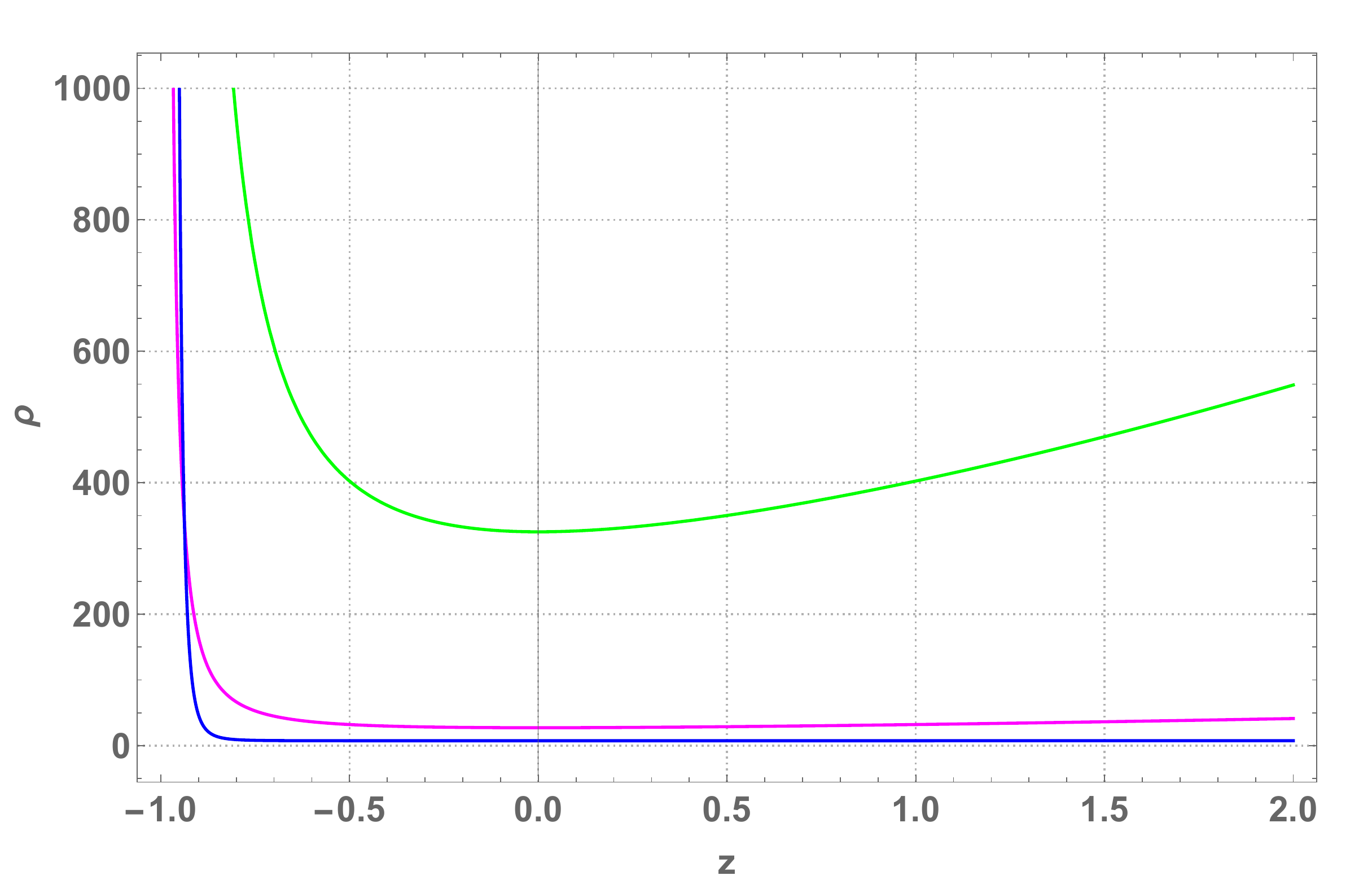}
\end{minipage}
\hfill
\begin{minipage}{55mm}
\includegraphics[width=60 mm]{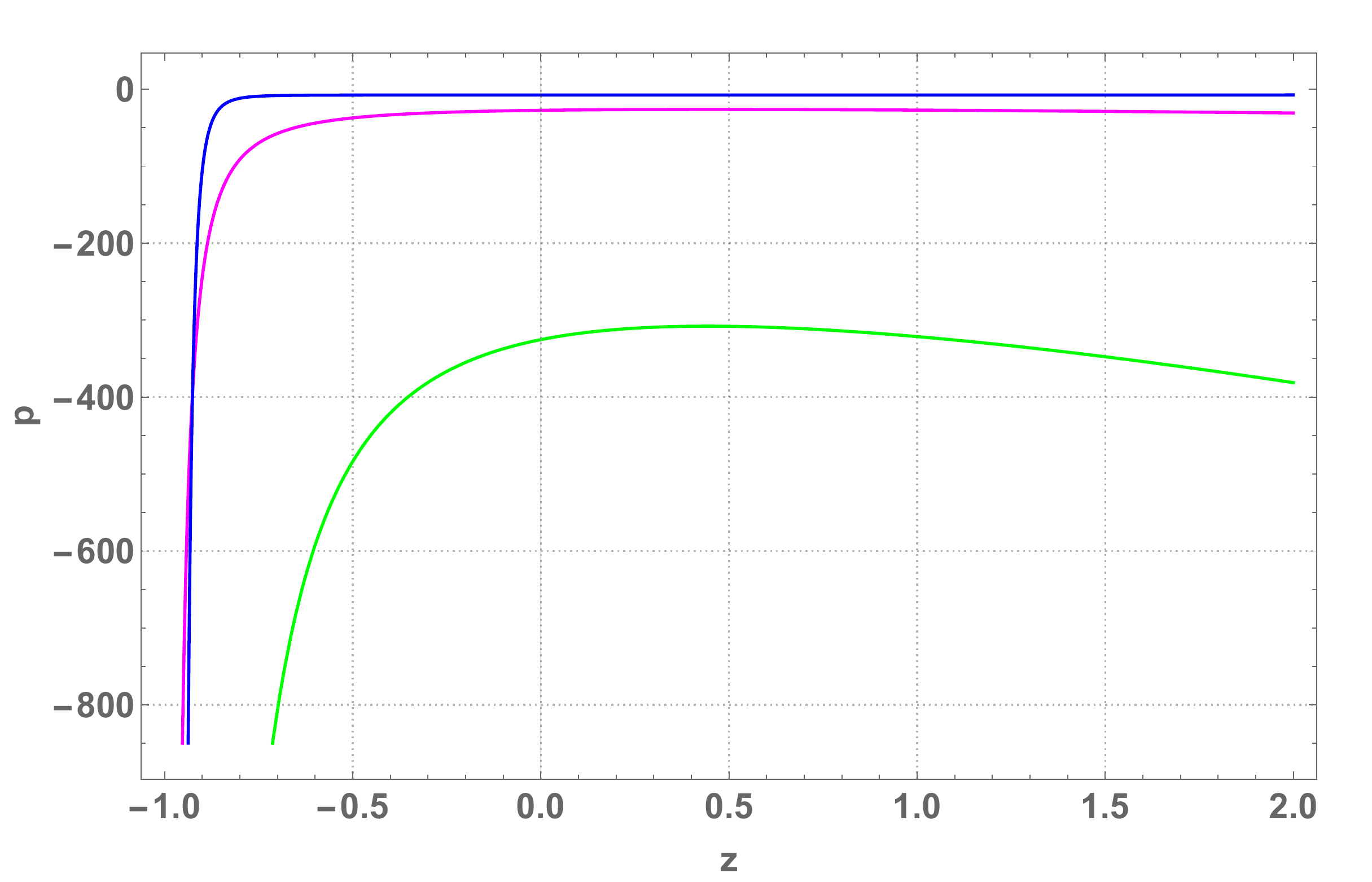}
\end{minipage}
\hfill
\begin{minipage}{55mm}
\includegraphics[width=60 mm]{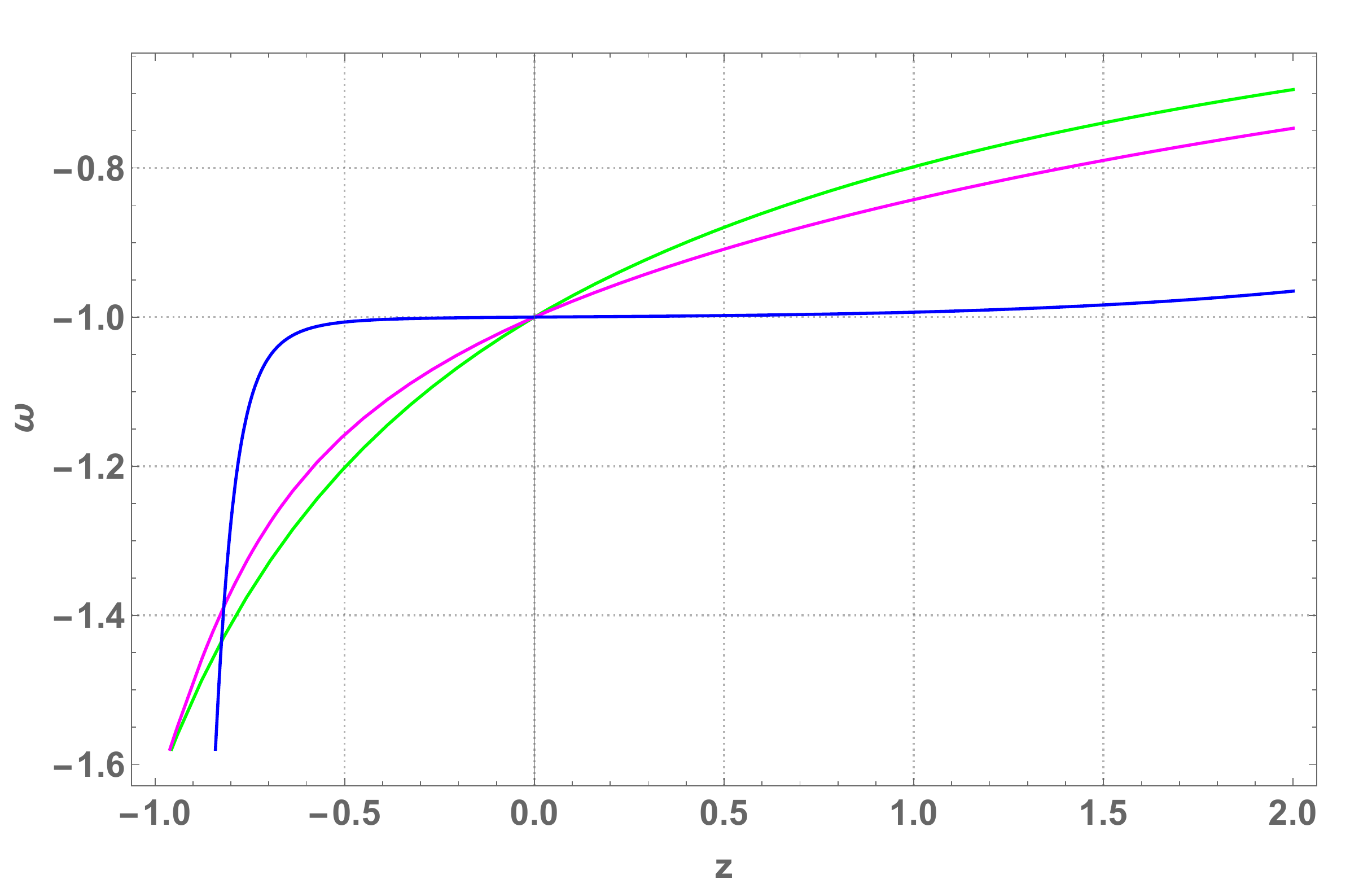}
\end{minipage}
\caption{Redshift evolution of $\rho$, $p$, and $\omega$ with $\alpha=15$, $\beta=-0.55$, $n=1$ for  ($m$, $k$)=($0.480012$, $0.5$)  [green curve], ($m$, $k$)= ($0.480003$, $0.25$) [magenta curve], and ($m$, $k$)=($1.15$, $0.1$)  [blue curve]. } \label{figcase2z}
\end{figure}

\textit{\textbf{Case--2: ($n=1$)}} The model in this case is in quadratic form, $f(Q)=\alpha Q^2+\beta$. The physical parameter behaviors for this choice with respect to time component and redshift parameter are depicted in the figure sets (\ref{figcase2t}--\ref{figcase2z}) for $\alpha=15$ and $\beta=-0.5$.

It can be observed that the same big rip singularity occurred as the linear case-1 at finite time $t=\frac{j\pi}{k}$, where $j$ is an integer (j=0,1,2,3.....). Here the positivity of density parameter depends on positive value of $\alpha$ as in linear case. Furthermore, the transitional periodic behavior of pressure with finite time singularity and EoS parameter from positive to negative phase  are obtained with the choice of ($m$, $k$)=($1.15$, $0.1$). The other two values of $m$ and $k$ lead to accelerating phase for pressure with singularity and for EoS parameter within the range $-1.7 <\omega < -0.3$ without singularity. \\   

\textit{\textbf{Case--3: ($n=-2$)}} 
In the above two cases, we had periodic evolution with finite time singularity both in accelerating phase and transitional phases. In order to address the periodic nature of universe in $f(Q)$ gravity background without any cosmic singularity, we admit the value of $n$ to be $-2$. The physical parameters in equation (\ref{rho-1}-\ref{omega-1}) with completely periodic behavior are depicted in the following figure sets for $\alpha=-5$ and $\beta=-15$.

\begin{figure}[H]
\centering
\begin{minipage}{55mm}
\includegraphics[width=60 mm]{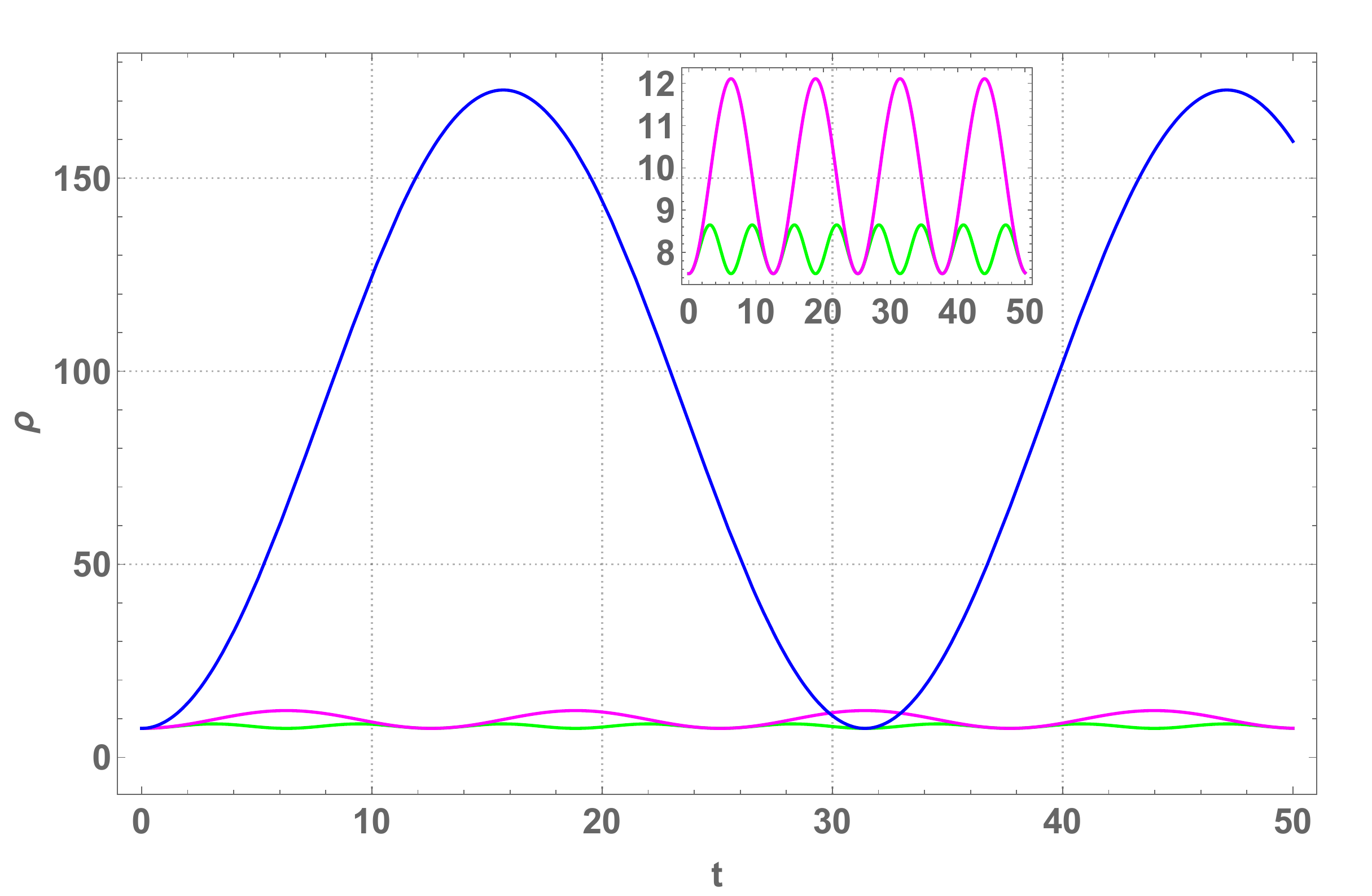}
\end{minipage}
\hfill
\begin{minipage}{55mm}
\includegraphics[width=60 mm]{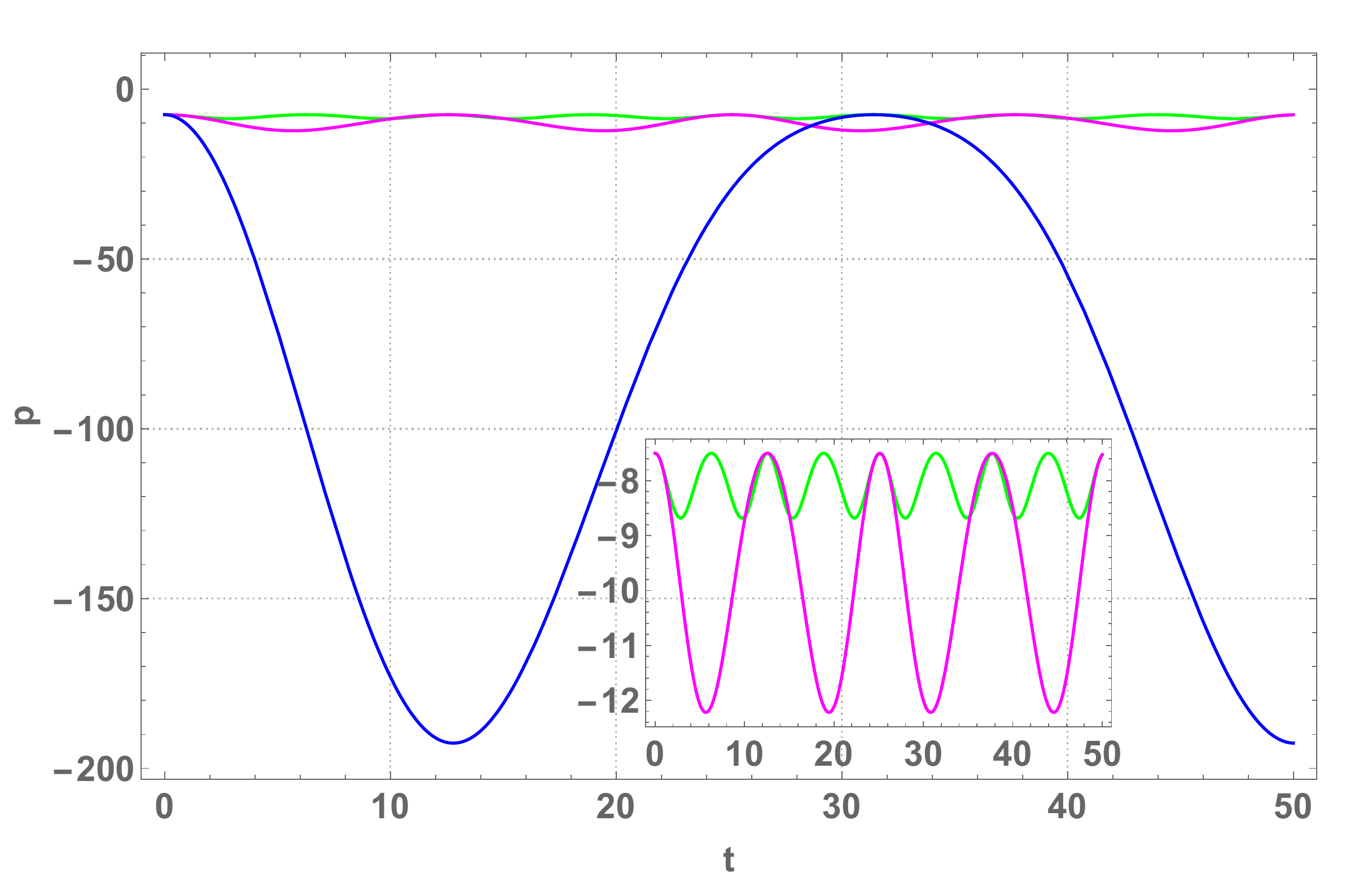}
\end{minipage}
\hfill
\begin{minipage}{55mm}
\includegraphics[width=60 mm]{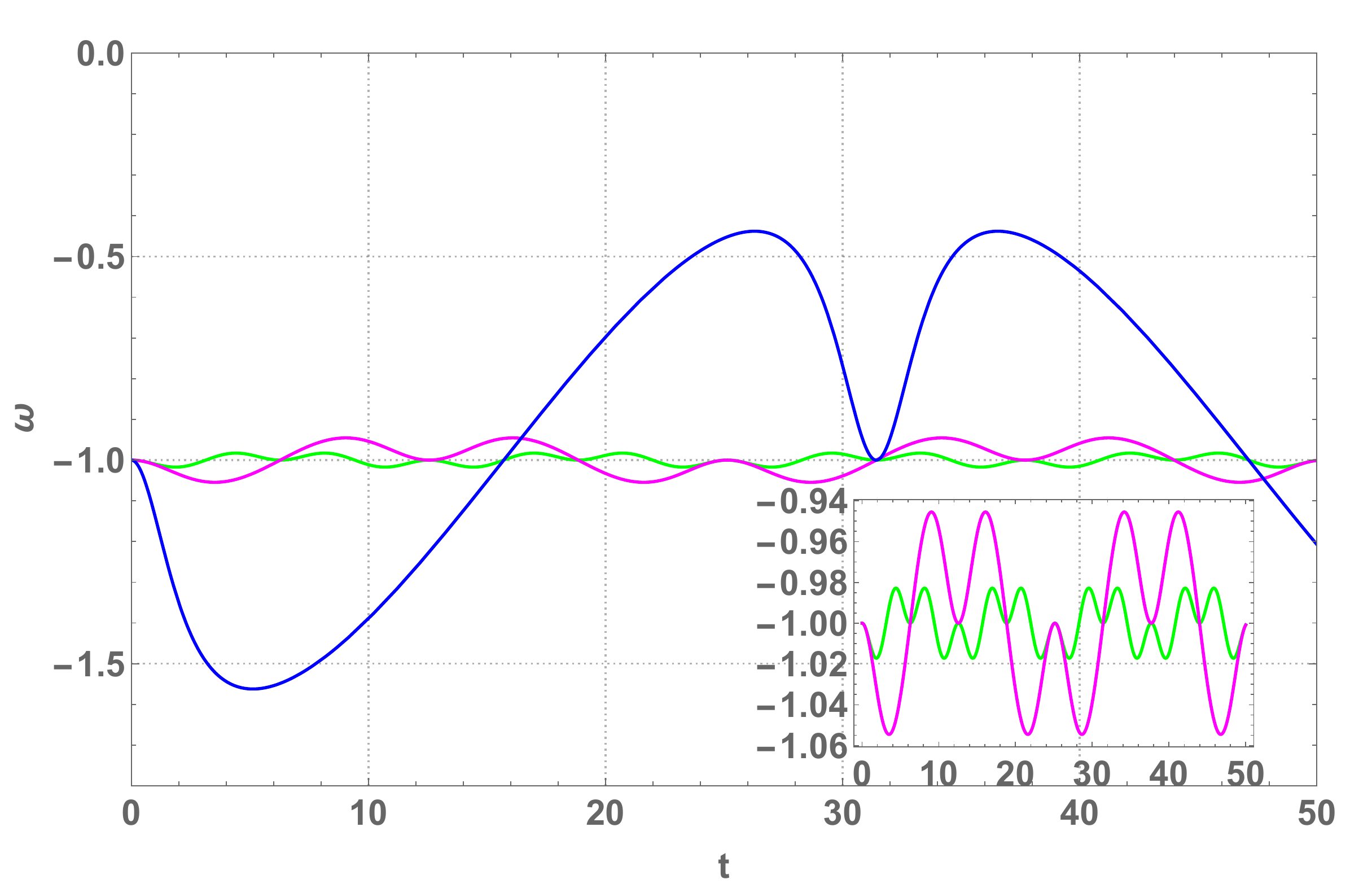}
\end{minipage}
\caption{Time (in $Gyr$) evolution of of $\rho$, $p$, and $\omega$ with $\alpha=-5$, $\beta=-15$,  $n=-2$, for  ($m$, $k$)=($0.480012$, $0.5$)  [green curve], ($m$, $k$)= ($0.480003$, $0.25$) [magenta curve], and ($m$, $k$)=($1.15$, $0.1$)  [blue curve].}\label{figcase3t}
\end{figure}

\begin{figure}[H]
\centering
\begin{minipage}{55mm}
\includegraphics[width=60 mm]{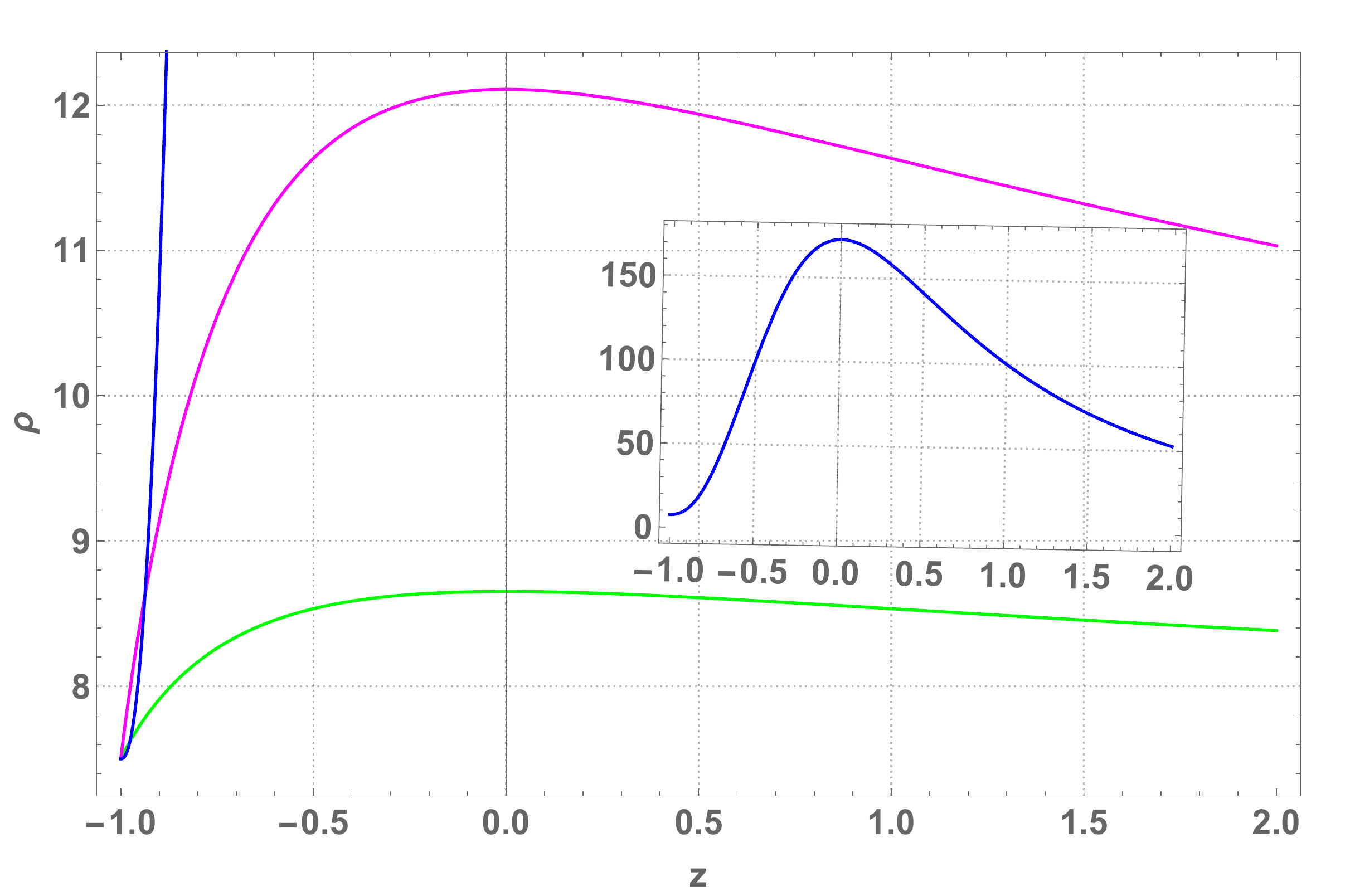}
\end{minipage}
\hfill
\begin{minipage}{55mm}
\includegraphics[width=60 mm]{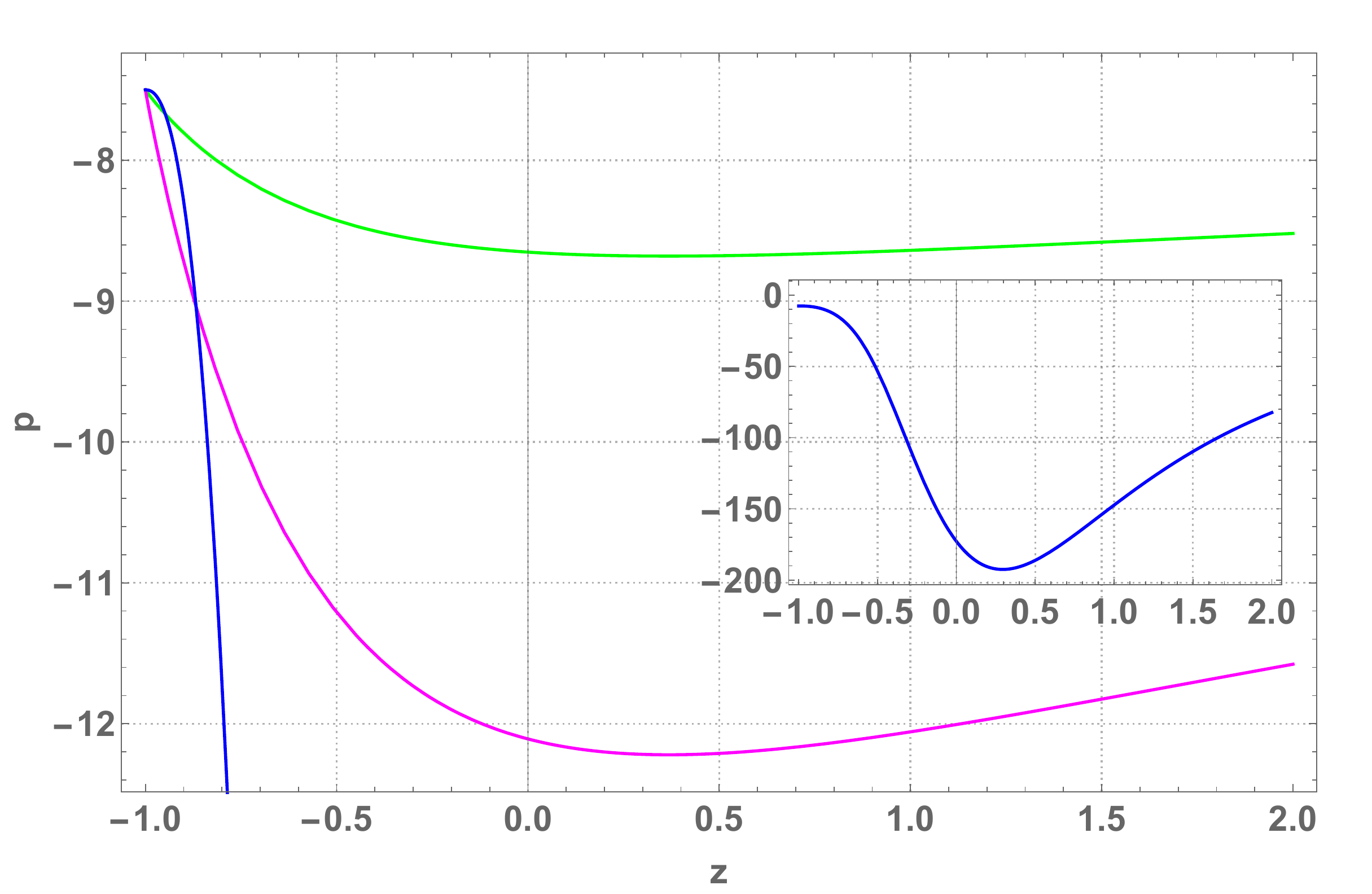}
\end{minipage}
\hfill
\begin{minipage}{55mm}
\includegraphics[width=60 mm]{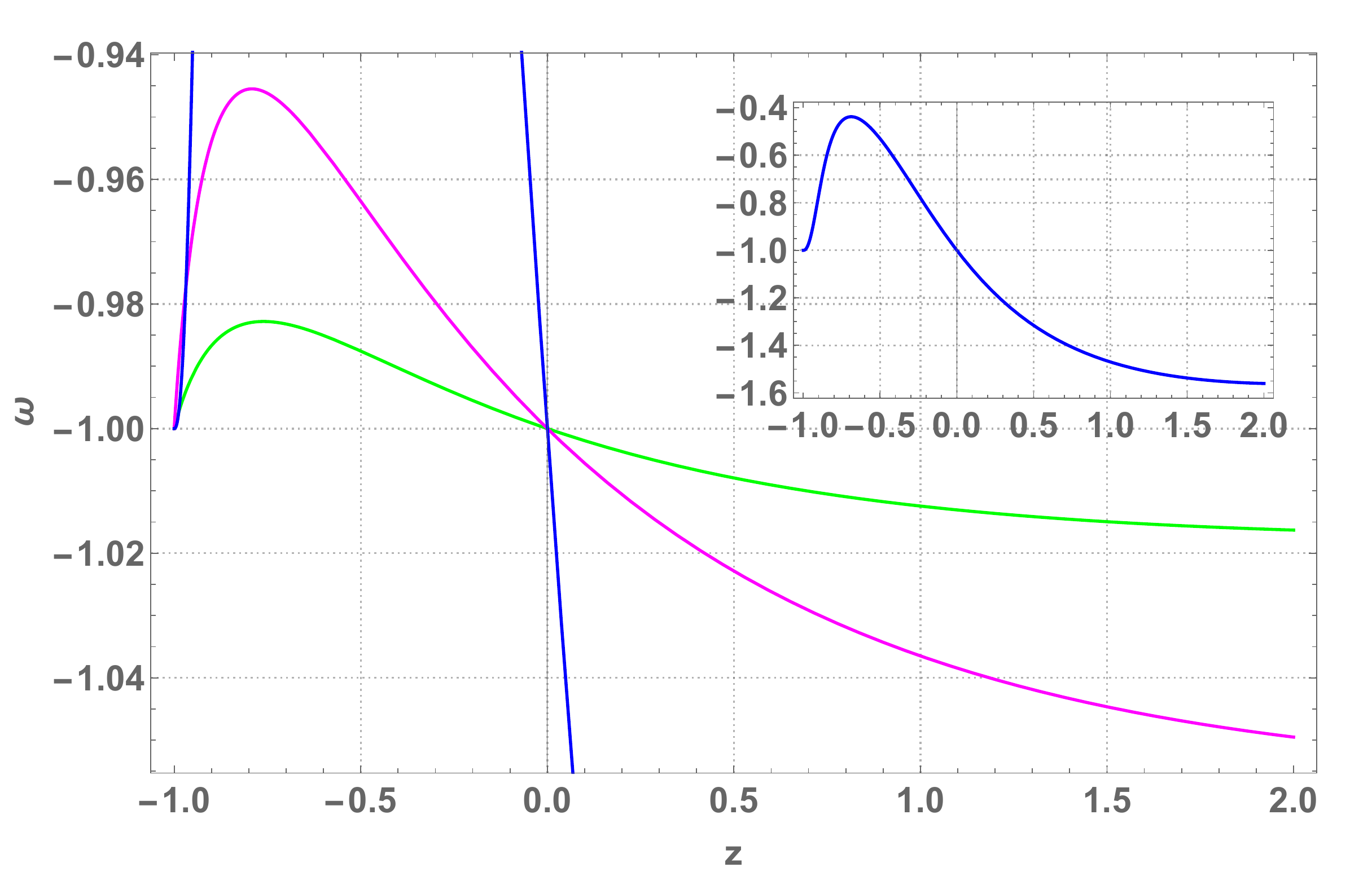}
 \label{figed-6}
\end{minipage}
\caption{Redshift evolution of $\rho$, $p$, and $\omega$ with $\alpha=-5$, $\beta=-15$,  $n=-2$, for  ($m$, $k$)=($0.480012$, $0.5$)  [green curve], ($m$, $k$)= ($0.480003$, $0.25$) [magenta curve], and ($m$, $k$)=($1.15$, $0.1$)  [blue curve]. }\label{figcase3z}
\end{figure}

In both figure profile (\ref{figcase3t}-\ref{figcase3z}), the periodicity occurred without singularity for each value of $m$ and $k$. The third set of values for $(m, k)$ provides cyclic evolution in high range of values, while the other two set of values are allowed for the periodic behavior in small range. The complete evolution for that are depicted in each plot. In case of the time evolution profile (\ref{figcase3t}), we show the enlarged view of ($m$, $k$)=($0.480012$, $0.5$), ($0.480003$, $0.25$) and in the redshift evolution profile (\ref{figcase3z}), the enlarged view of ($m$, $k$)=($1.15$, $0.1$) is shown to confirm the cyclic nature without any singularity.

\section{Energy Conditions (ECs)}\label{sec5}
It is well-known that the ECs represent the attractive nature of gravity, besides assigning the fundamental causal and the geodesic structure of spacetime \cite{E}. The different models of $f(Q)$ gravity give rise to the problem of viability. By imposing the ECs, we may have some constrains on the $f(Q)$ model parameters \cite{Gb}. The different ECs are used to obtain the dynamics of the solutions for a plenty of problems. For example the strong energy condition (SEC): $\rho+3p \geq 0$, and weak energy condition (WEC): $\rho+p\ge0$ and $\rho\ge 0$ were used in the Hawking-Penrose singularity theorems and the null energy condition (NEC): $\rho+p\ge 0$ is required in order to prove the second law of black hole thermodynamics. The ECs were primarily formulated in GR \cite{H}, and later derived in several modified gravity theories by introducing effective pressure and energy density. Similarly, the dominant energy condition (DEC): $\rho-p\ge 0$ is considered for stabilizing a model as it limited the velocity of energy transfer to the speed of light. That means the  mass-energy can never be observed to be flowing faster than light. In the present study, the corresponding ECs are derived as 
\begin{equation}
 {\rho+p}=\alpha  m 6^n (n+1) (2 n+1) \sin (2 k t) \csc (k t) \left(\frac{k^2 \csc ^2(k t)}{m^2}\right)^{n+1}\ge 0,
\end{equation}

\begin{equation}
{\rho-p}=\alpha  \left(-2^{n+1}\right) 3^n (2 n+1) (m (n+1) \cos (k t)-3) \left(\frac{k^2 \csc ^2(k t)}{m^2}\right)^{n+1}-\beta\ge 0,
\end{equation}

\begin{equation}
 {\rho+3p}=\frac{\alpha  m^2 6^{n+1} (2 n+1) (m (n+1) \cos (k t)-1) \left(\frac{k^2 \csc ^2(k t)}{m^2}\right)^{n+1}+\beta  m^2}{m^2}\ge 0.
\end{equation}

\begin{figure}[H]
\centering
\begin{minipage}{55mm}
\includegraphics[width=60 mm]{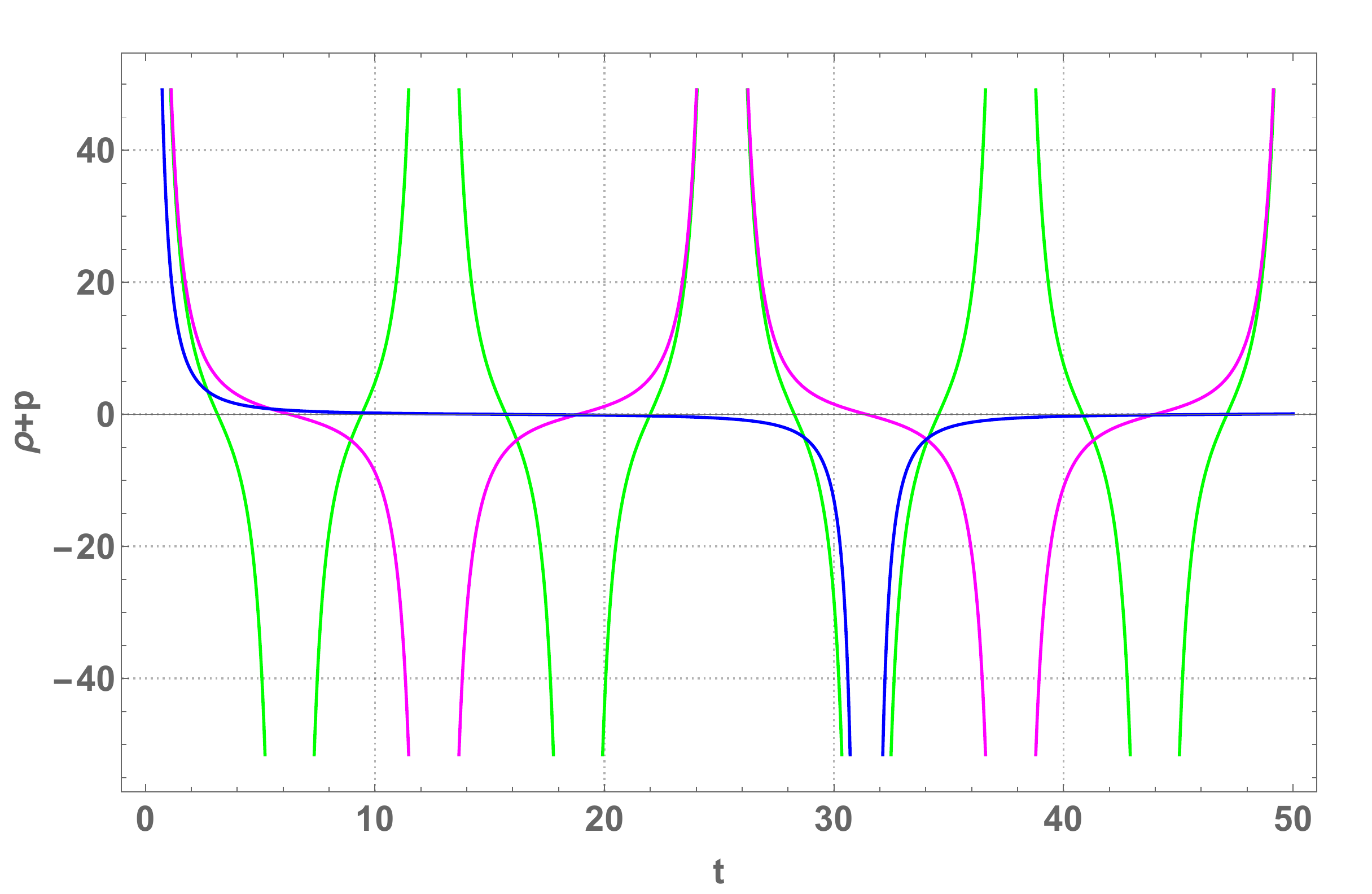}
\end{minipage}
\hfill
\begin{minipage}{55mm}
\includegraphics[width=60 mm]{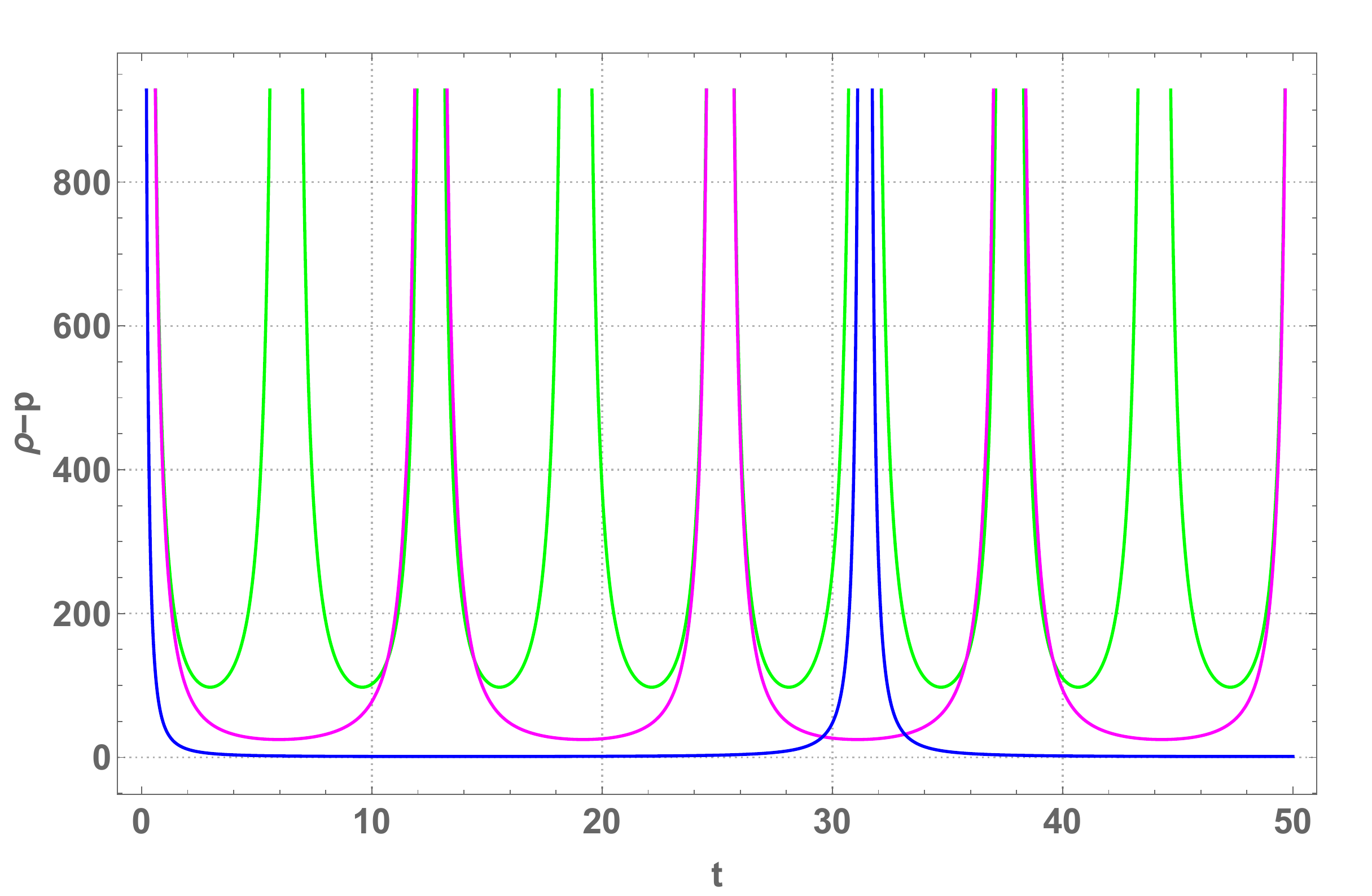}
\end{minipage}
\hfill
\begin{minipage}{55mm}
\includegraphics[width=60 mm]{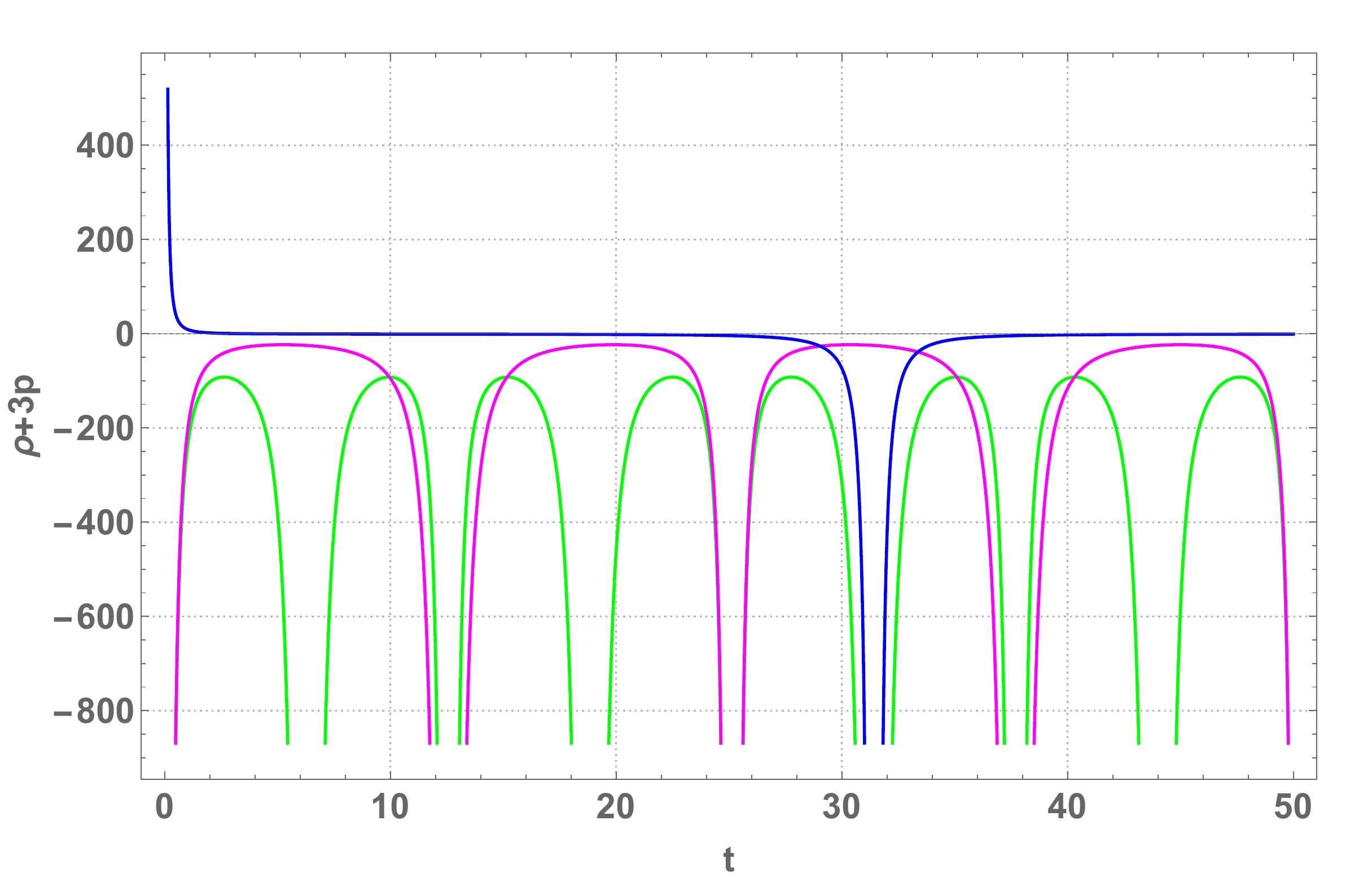}
 \label{figec-5}
\end{minipage}
\caption{Time (in $Gyr$) evolution of ECs with $\alpha=15$, $\beta=-0.5$, $n=0$ for ($m$, $k$)=($0.480012$, $0.5$)  [green curve], ($m$, $k$)= ($0.480003$, $0.25$) [magenta curve], and ($m$, $k$)=($1.15$, $0.1$)  [blue curve].}\label{figeccase1t}
\end{figure}

\begin{figure}[H]
\centering
\begin{minipage}{55mm}
\includegraphics[width=60 mm]{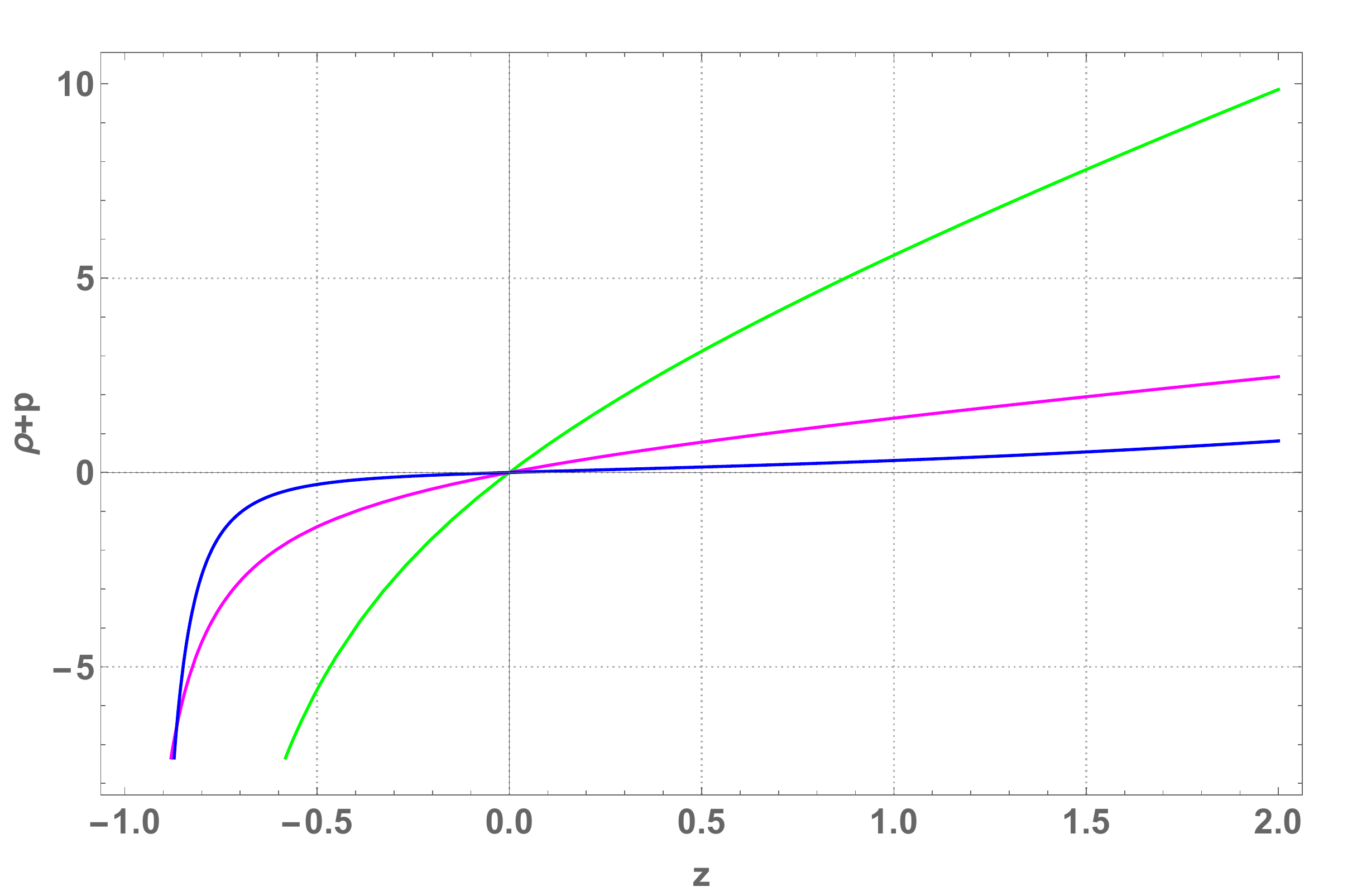}
\end{minipage}
\hfill
\begin{minipage}{55mm}
\includegraphics[width=60 mm]{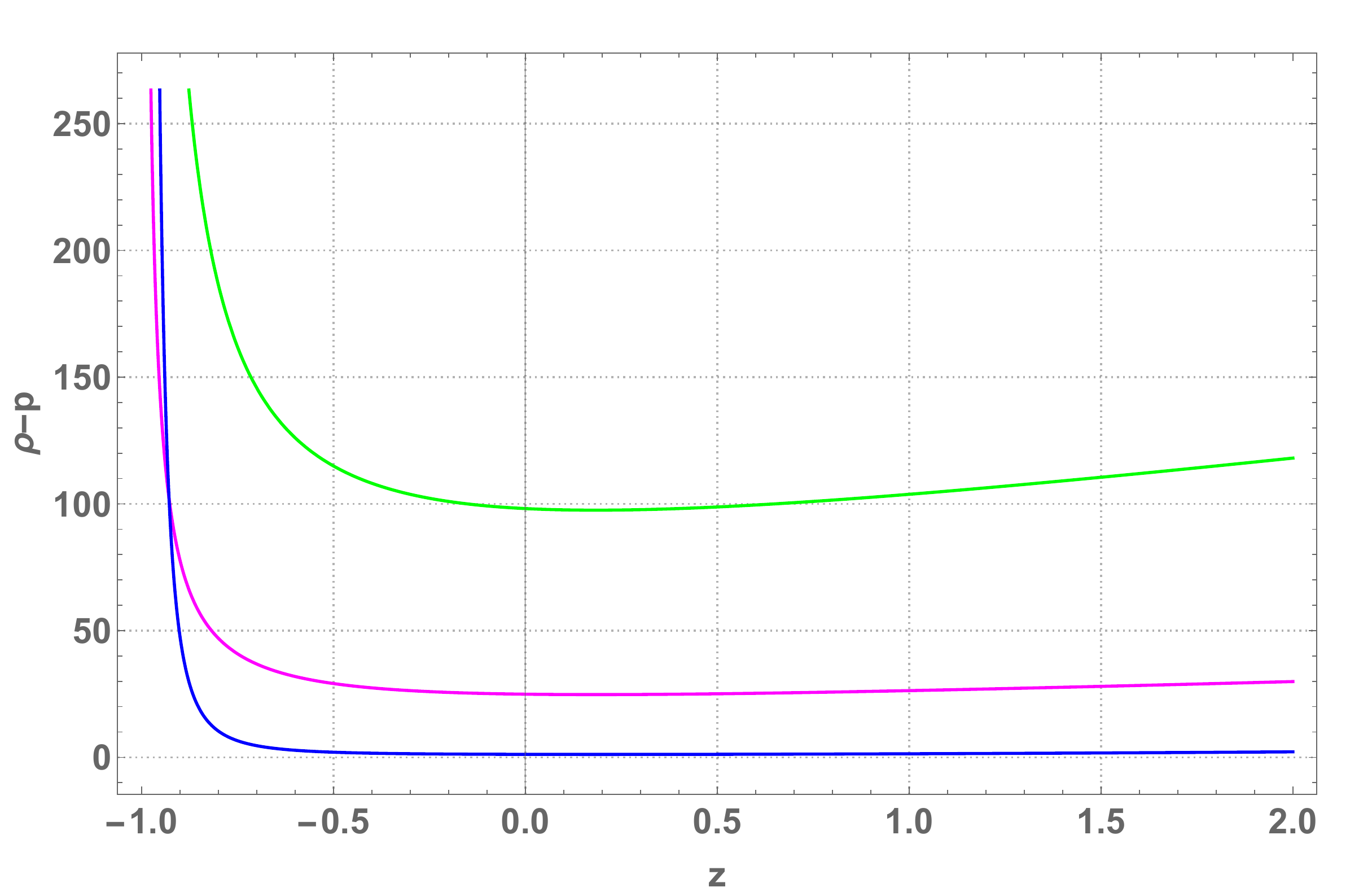}
\end{minipage}
\hfill
\begin{minipage}{55 mm}
\includegraphics[width=60 mm]{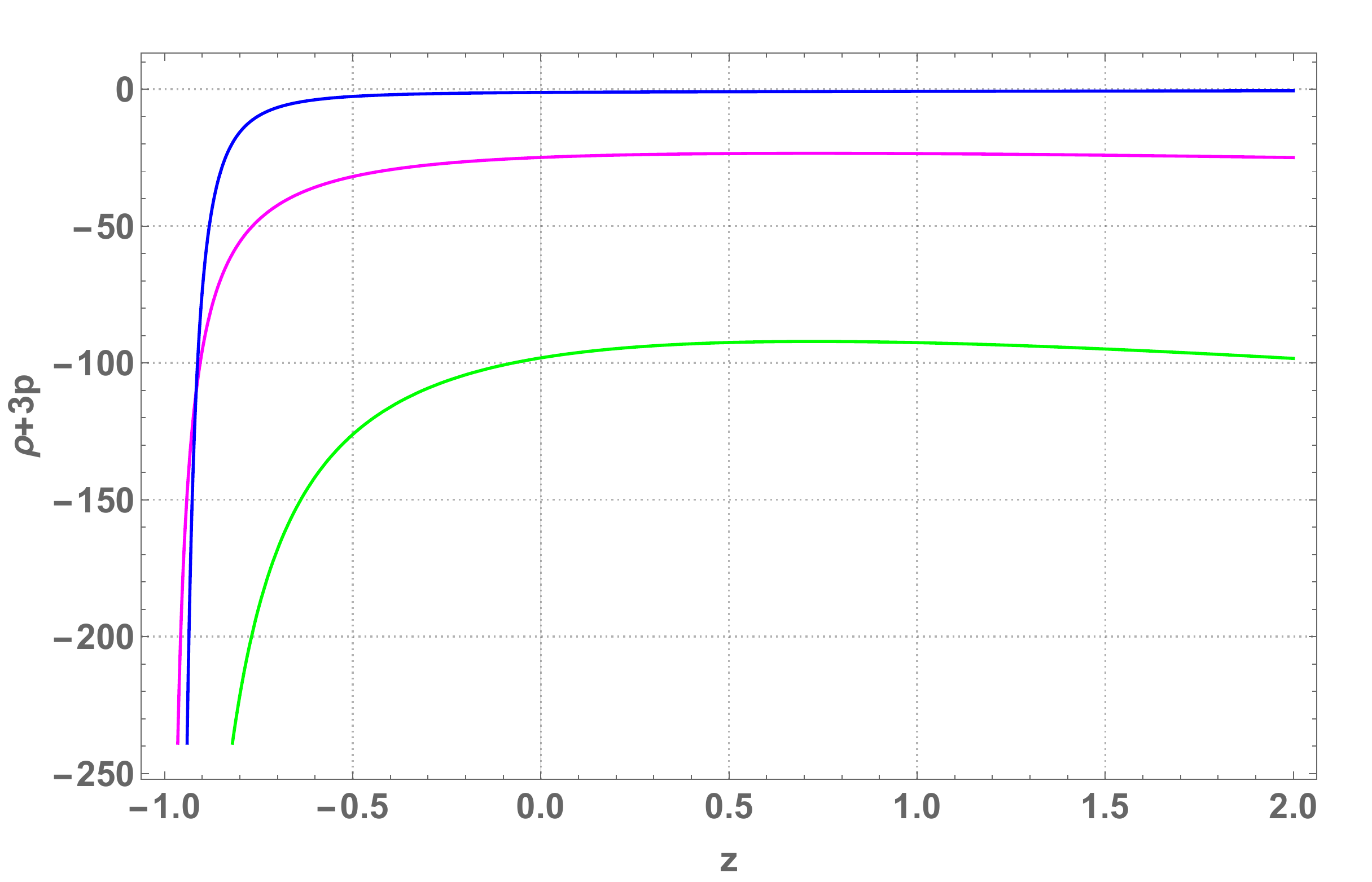}
\end{minipage}
\caption{Redshift evolution of ECs with $\alpha=15$, $\beta=-0.5$, $n=0$ for ($m$, $k$)=($0.480012$, $0.5$)  [green curve], ($m$, $k$)= ($0.480003$, $0.25$) [magenta curve], and ($m$, $k$)=($1.15$, $0.1$)  [blue curve].}\label{figeccase1z}
\end{figure}

\begin{figure}[H]
\centering
\begin{minipage}{55mm}
\includegraphics[width=60 mm]{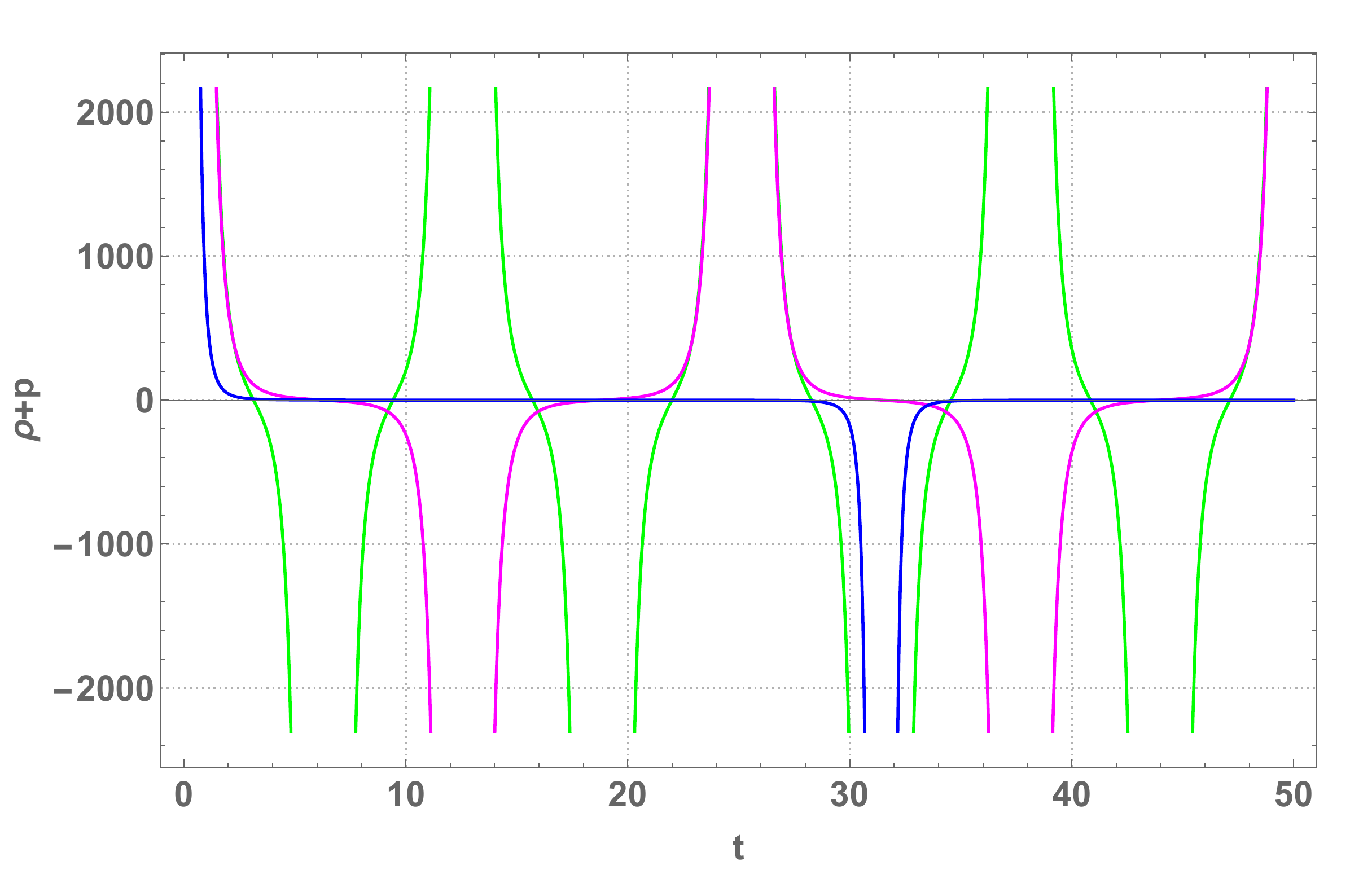}
\end{minipage}
\hfill
\begin{minipage}{55mm}
\includegraphics[width=60 mm]{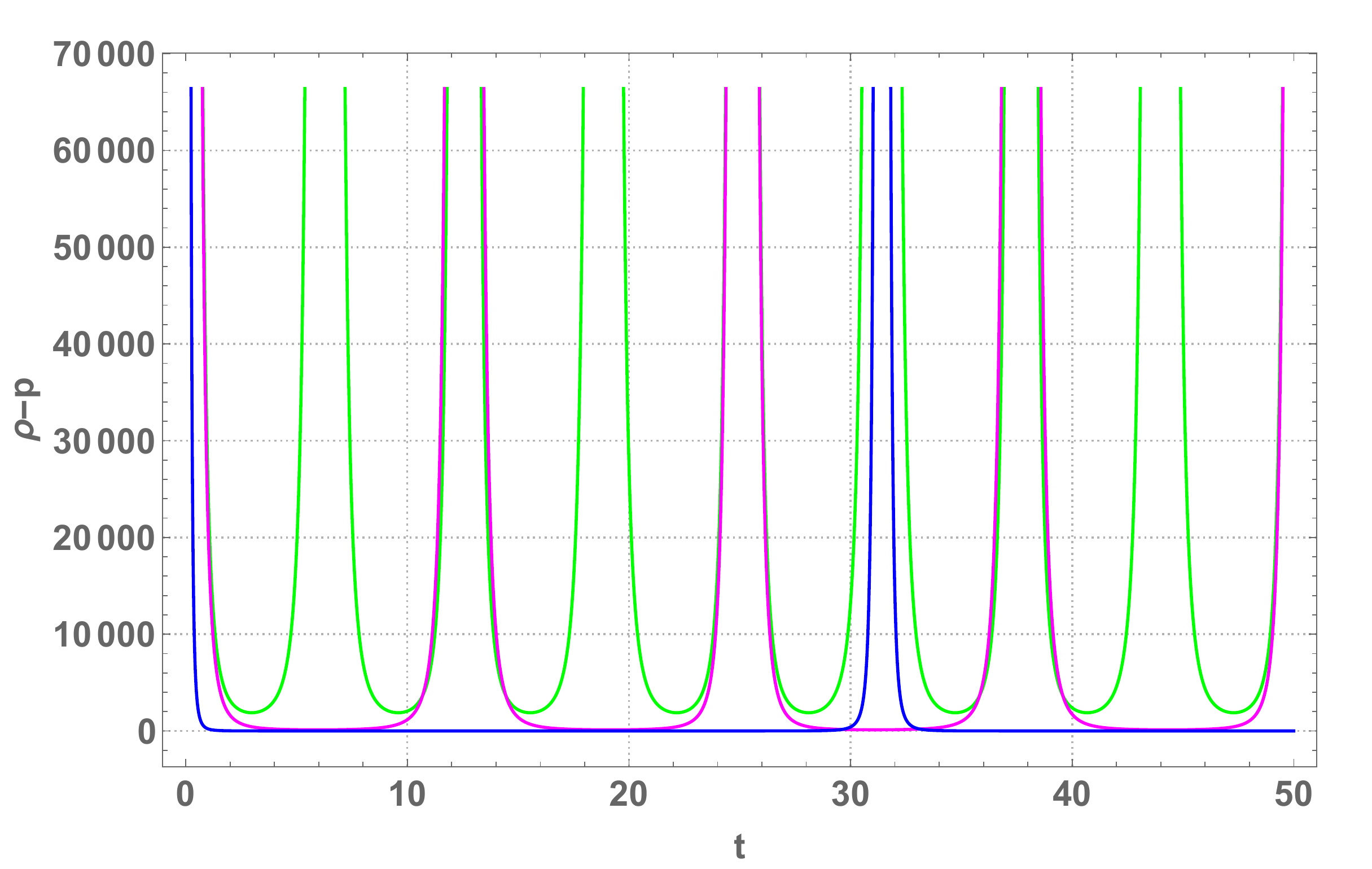}
\end{minipage}
\hfill
\begin{minipage}{55mm}
\includegraphics[width=60 mm]{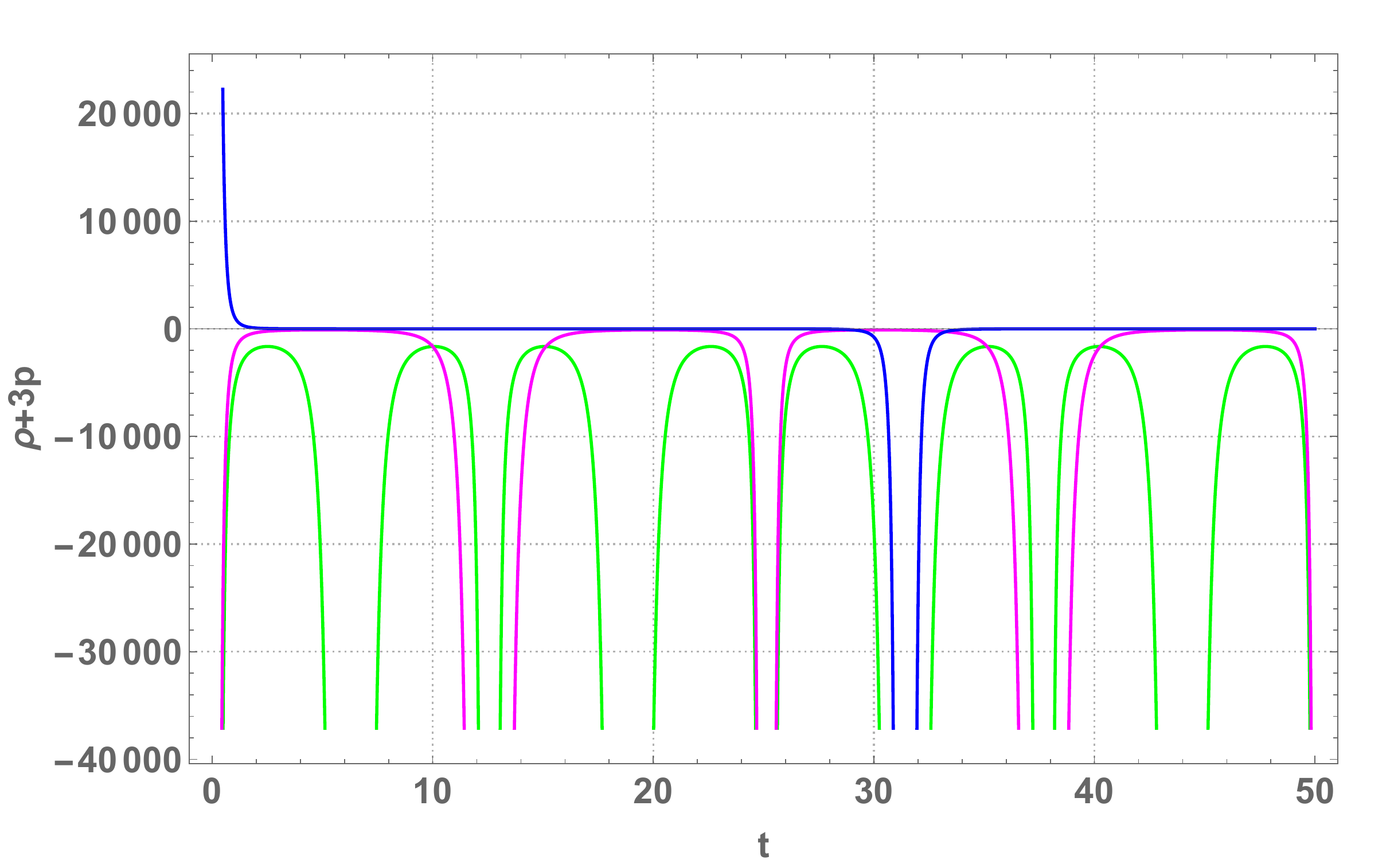}
\end{minipage}
\caption{Time (in $Gyr$) evolution of ECs with $\alpha=15$, $\beta=-0.5$, $n=1$ for ($m$, $k$)=($0.480012$, $0.5$)  [green curve], ($m$, $k$)= ($0.480003$, $0.25$) [magenta curve], and ($m$, $k$)=($1.15$, $0.1$)  [blue curve].}\label{figeccase2t}
\end{figure}

\begin{figure}[H]
\centering
\begin{minipage}{55mm}
\includegraphics[width=60 mm]{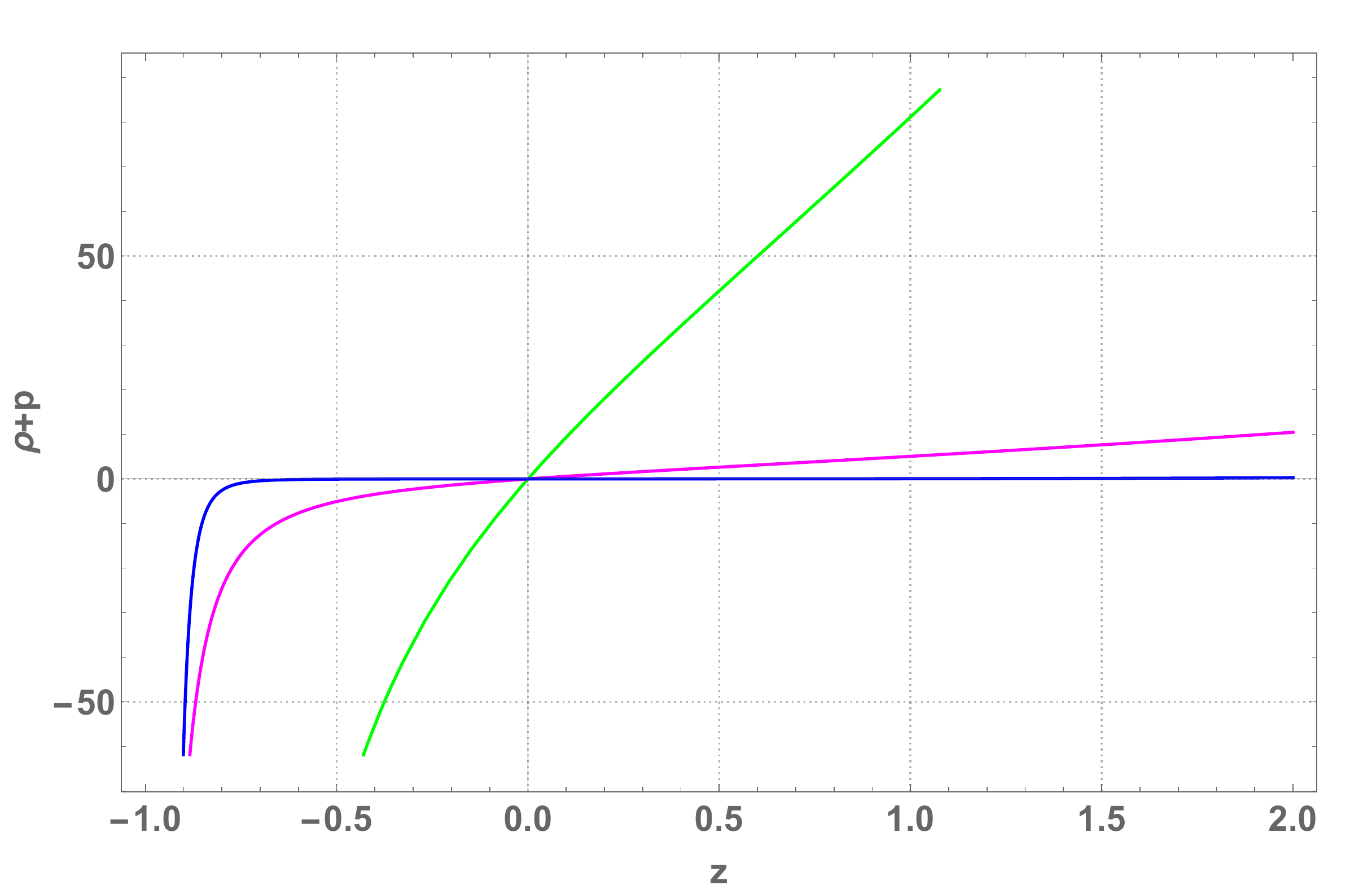}
\end{minipage}
\hfill
\begin{minipage}{55mm}
\includegraphics[width=60 mm]{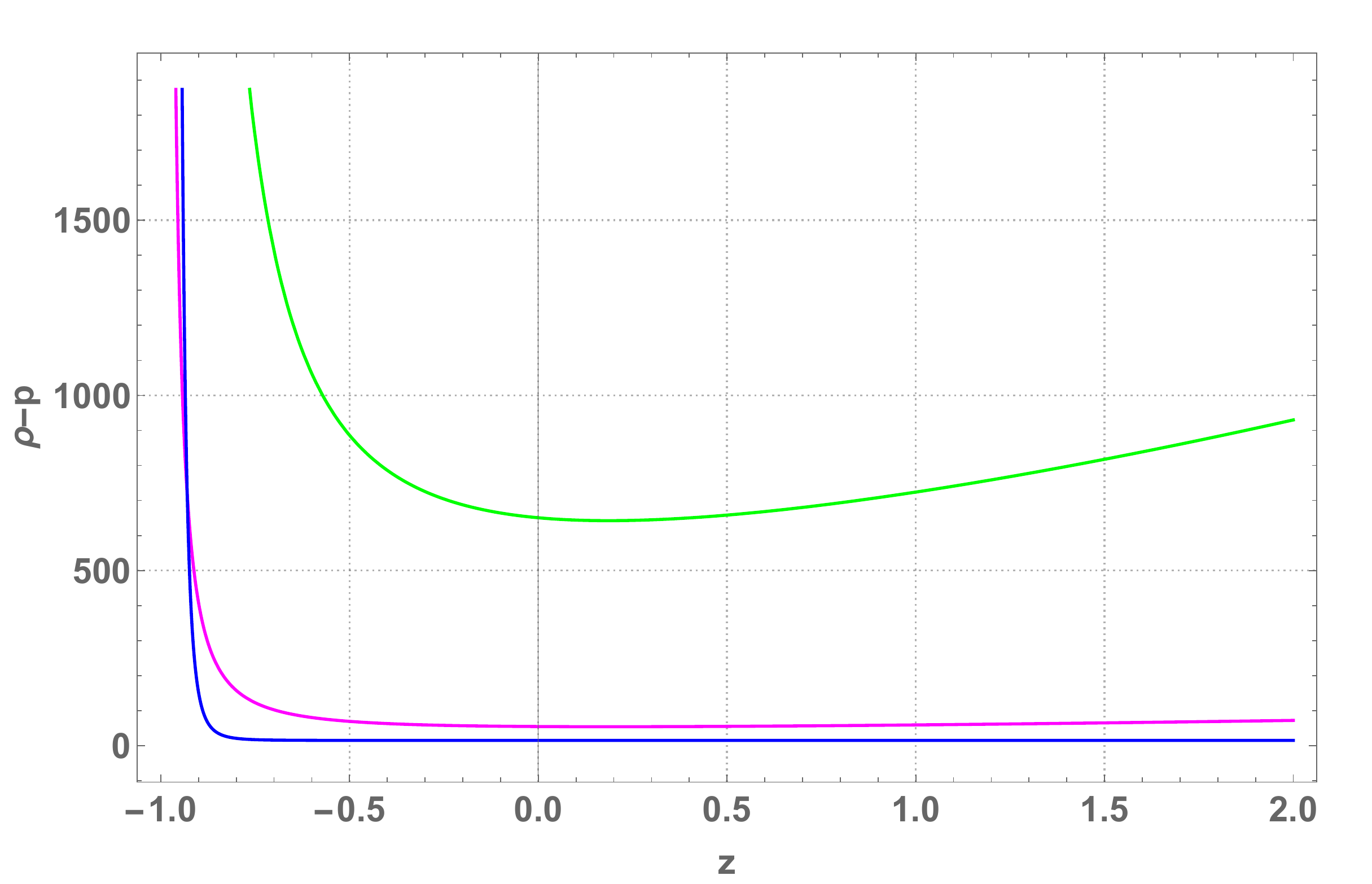}
\end{minipage}
\hfill
\begin{minipage}{55 mm}
\includegraphics[width=60 mm]{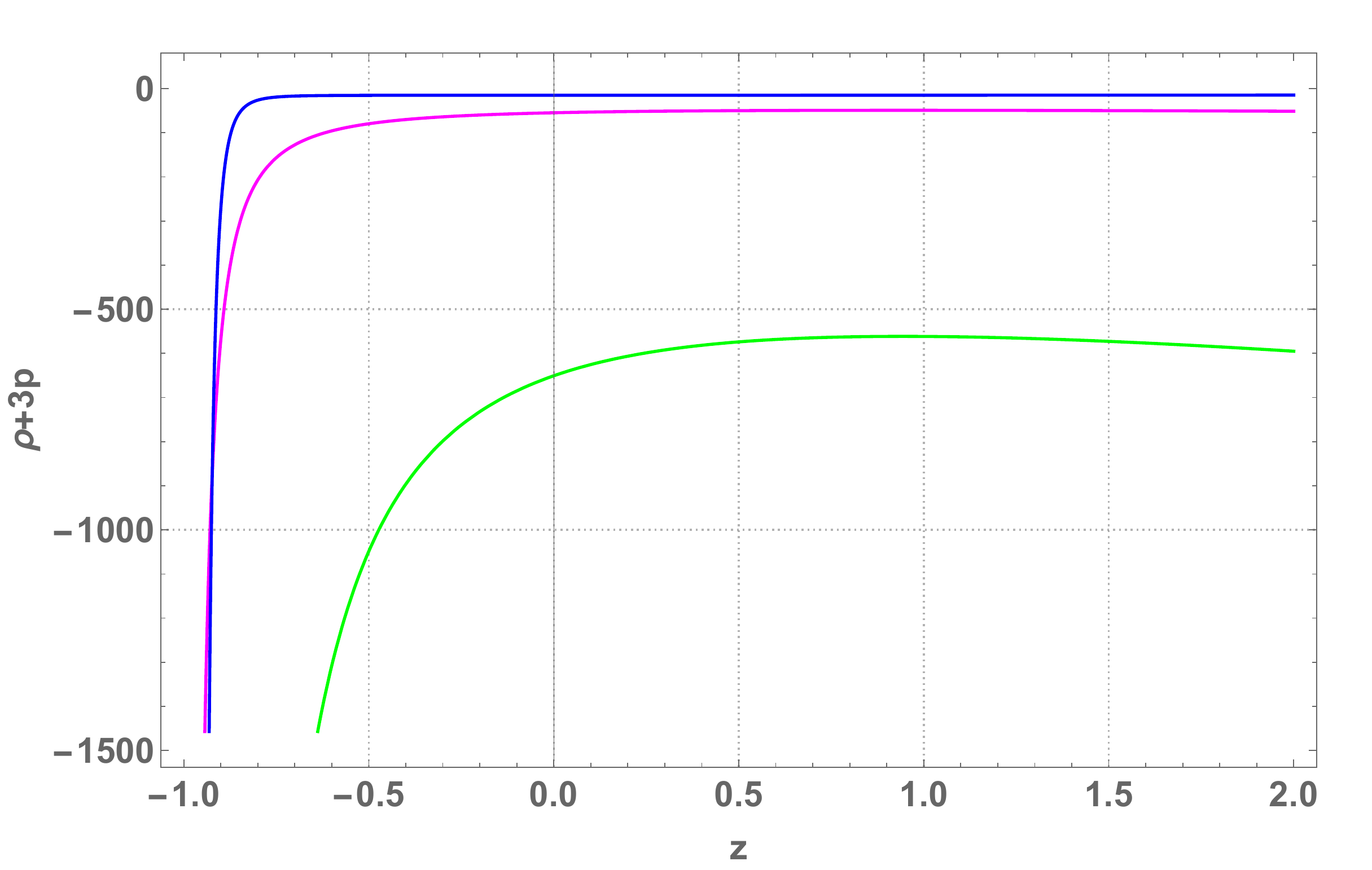}
\end{minipage}
\caption{Redshift evolution of ECs with $\alpha=15$, $\beta=-0.5$, $n=1$ for ($m$, $k$)=($0.480012$, $0.5$)  [green curve], ($m$, $k$)= ($0.480003$, $0.25$) [magenta curve], and ($m$, $k$)=($1.15$, $0.1$)  [blue curve].}\label{figeccase2z}
\end{figure}
In the figure profile (\ref{figeccase1t}-\ref{figeccase2z}), the ECs for the linear and quadratic choice of $f(Q)$ gravity are depicted in details for $\alpha >0$ and $\beta <0$. The singularity appears in each plot at constant finite time which repeats periodically. NEC is satisfied at an initial phase and then violated for each $m$ and $k$, whereas SEC is only satisfied initially for ($m$, $k$)=($1.15$, $0.1$) and violated for other values and so on. At the time of Phantom era, the big rip singularity arises at finite time and it repeats periodically. From the evolutionary behavior of EoS parameter in figure (\ref{figcase1t} and \ref{figcase2t}), it can be observed that the ECs are satisfied when the universe lies in non-phantom era. Later on they are violated when the universe crossed the phantom line and entered the phantom phase. The presence of phantom fluid yields the NEC violation along with SEC as well. The DEC is satisfied in both approaches to stabilize the model.

\begin{figure}[H]
\centering
\begin{minipage}{55mm}
\includegraphics[width=60 mm]{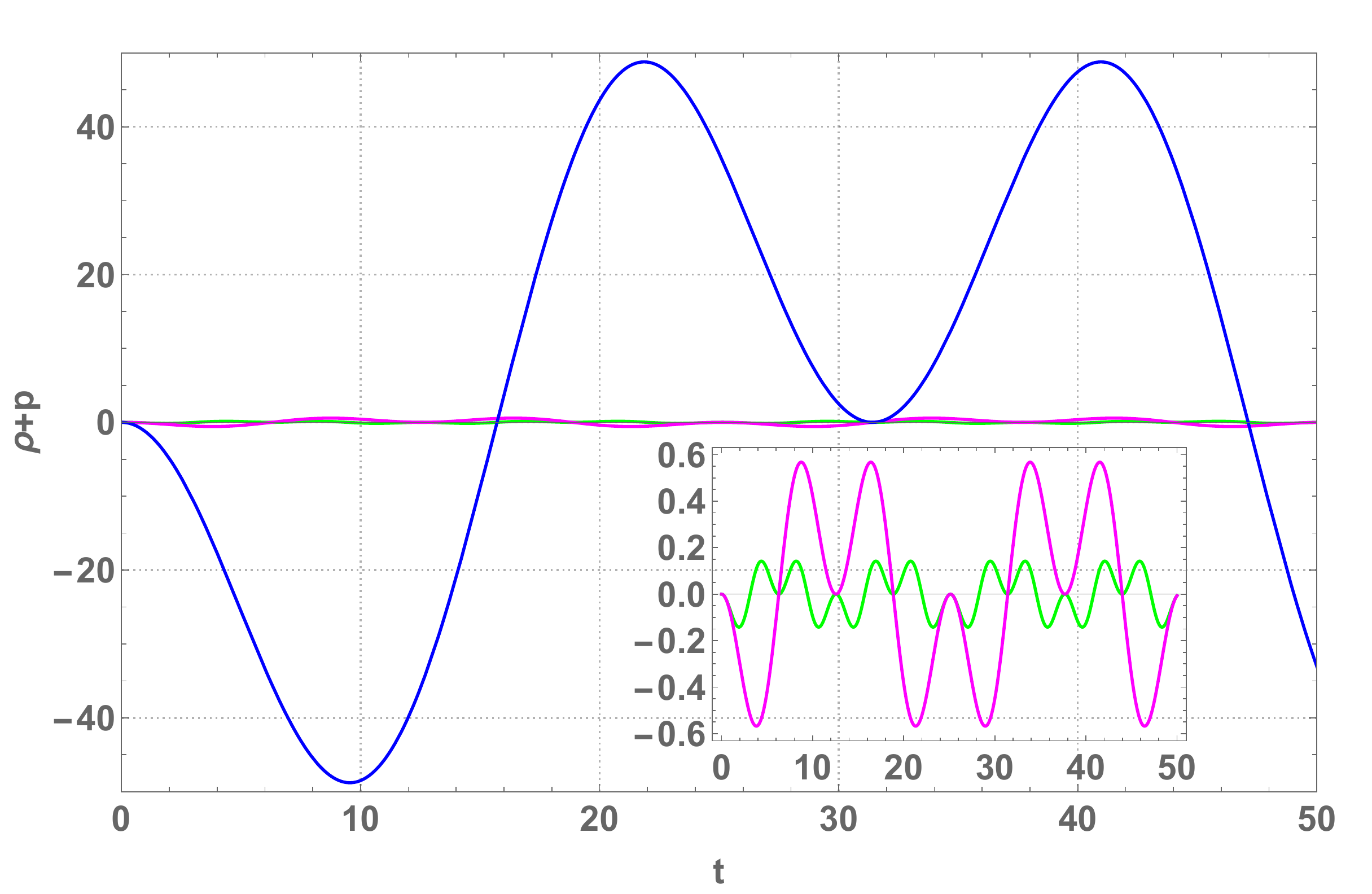}
\end{minipage}
\hfill
\begin{minipage}{55mm}
\includegraphics[width=60 mm]{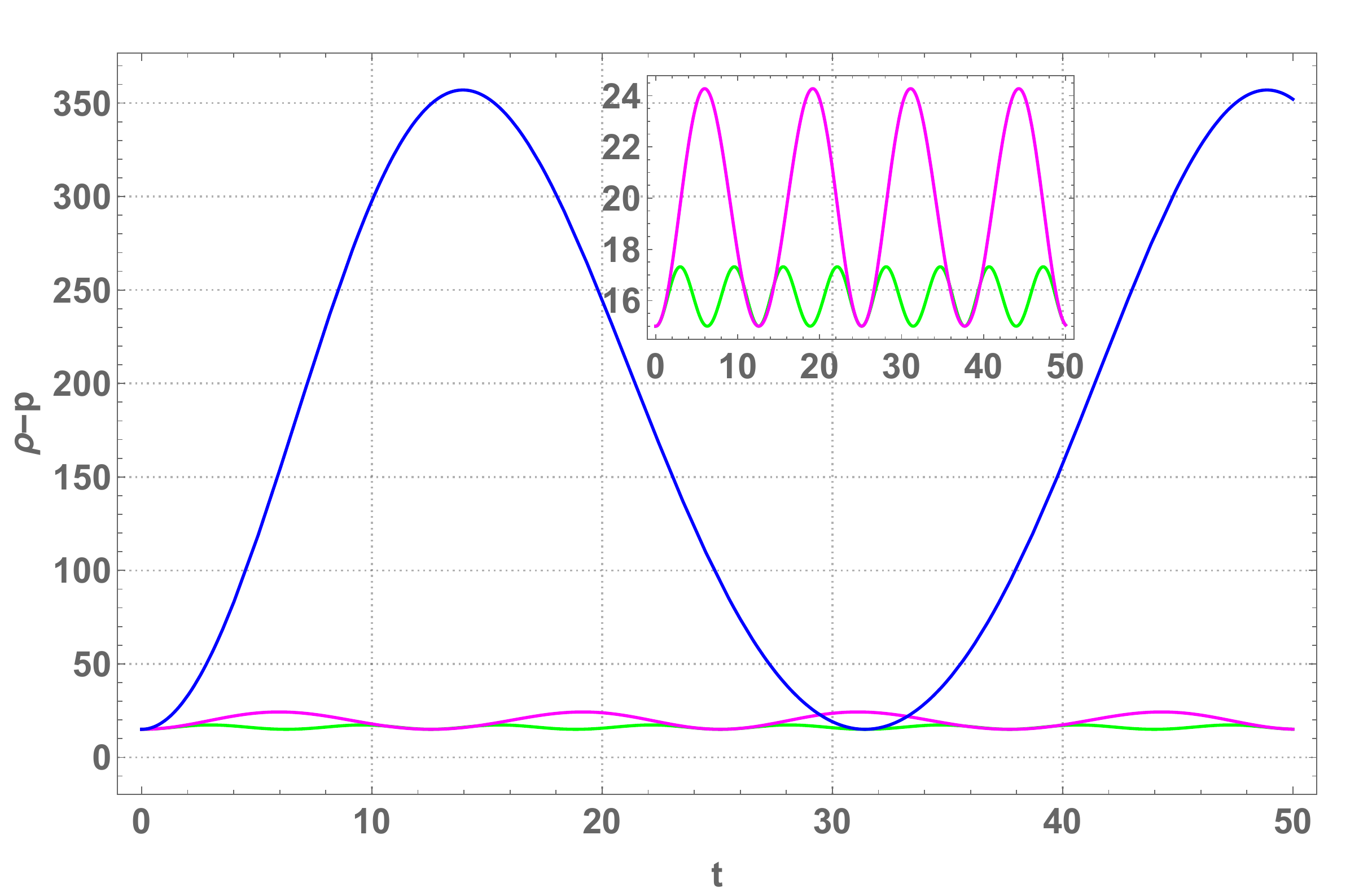}
\end{minipage}
\hfill
\begin{minipage}{55mm}
\includegraphics[width=60 mm]{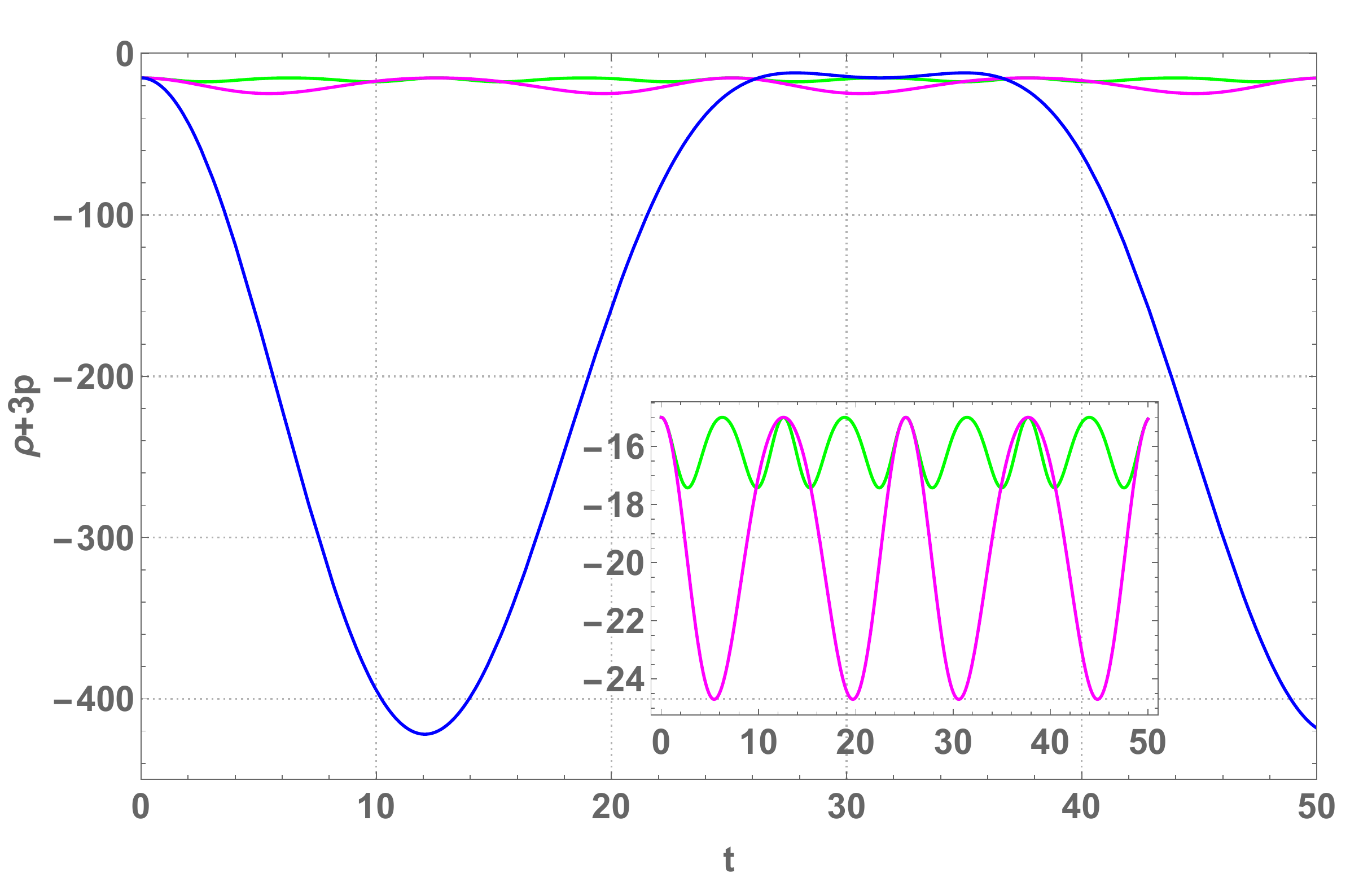}
\end{minipage}
\caption{Time (in $Gyr$) evolution of ECs with $\alpha=-5$, $\beta=-15$,  $n=-2$, for  ($m$, $k$)=($0.480012$, $0.5$)  [green curve], ($m$, $k$)= ($0.480003$, $0.25$) [magenta curve], and ($m$, $k$)=($1.15$, $0.1$)  [blue curve]. }\label{figeccase3t}
\end{figure}

  For the third case we observe the cyclic behavior in all ECs with the specific choice of $f(Q)$ gravity functional with $\alpha <0$ and $\beta <0$. The given choice of $f(Q)$ function in this model provide a cyclic universe without any singularity. At an initial phantom phase of universe, we were experiencing the violation of NEC and SEC. Later on after entering to non-phantom era, all ECs are satisfied and repeats it periodically without any singularity. However, the DEC is satisfied in all aspects. The complete evolution for this case is depicted in the following plot profile (\ref{figeccase3t}-\ref{figeccase3z}),      

\begin{figure}[H]
\centering
\begin{minipage}{55 mm}
\includegraphics[width=60 mm]{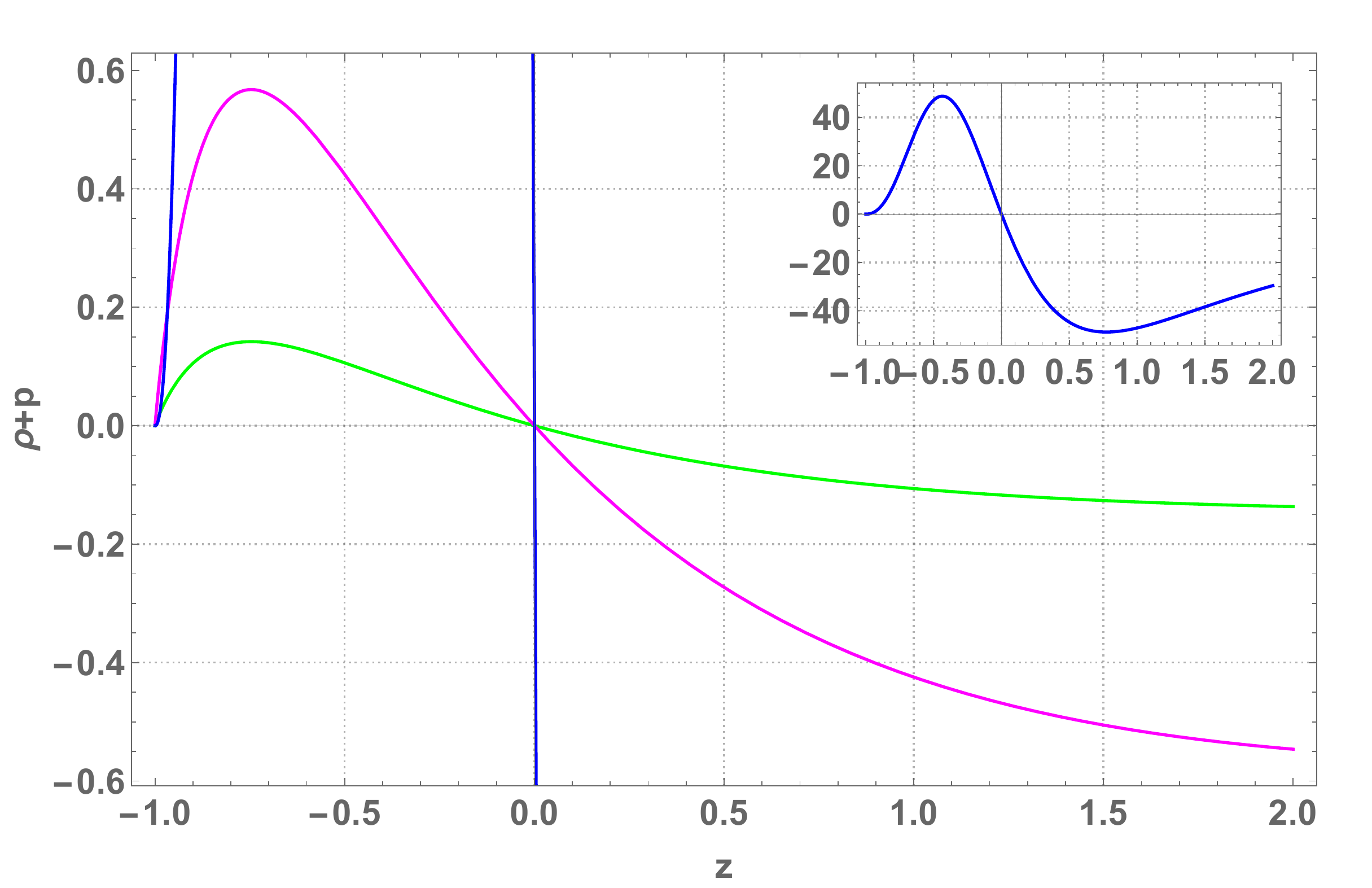}
\end{minipage}
\hfill
\begin{minipage}{55 mm}
\includegraphics[width=60 mm]{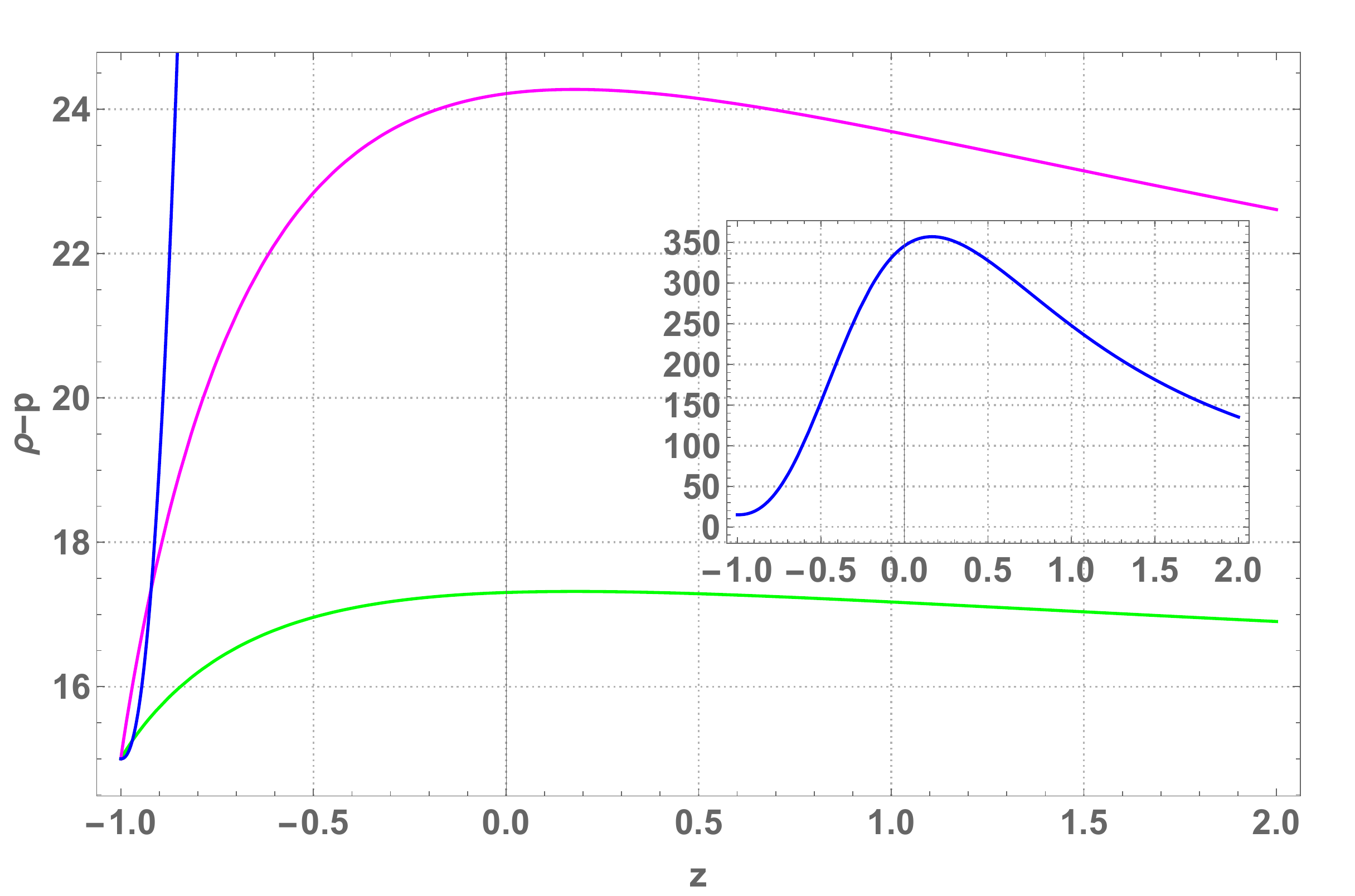}
\end{minipage}
\hfill
\begin{minipage}{55 mm}
\includegraphics[width=62 mm]{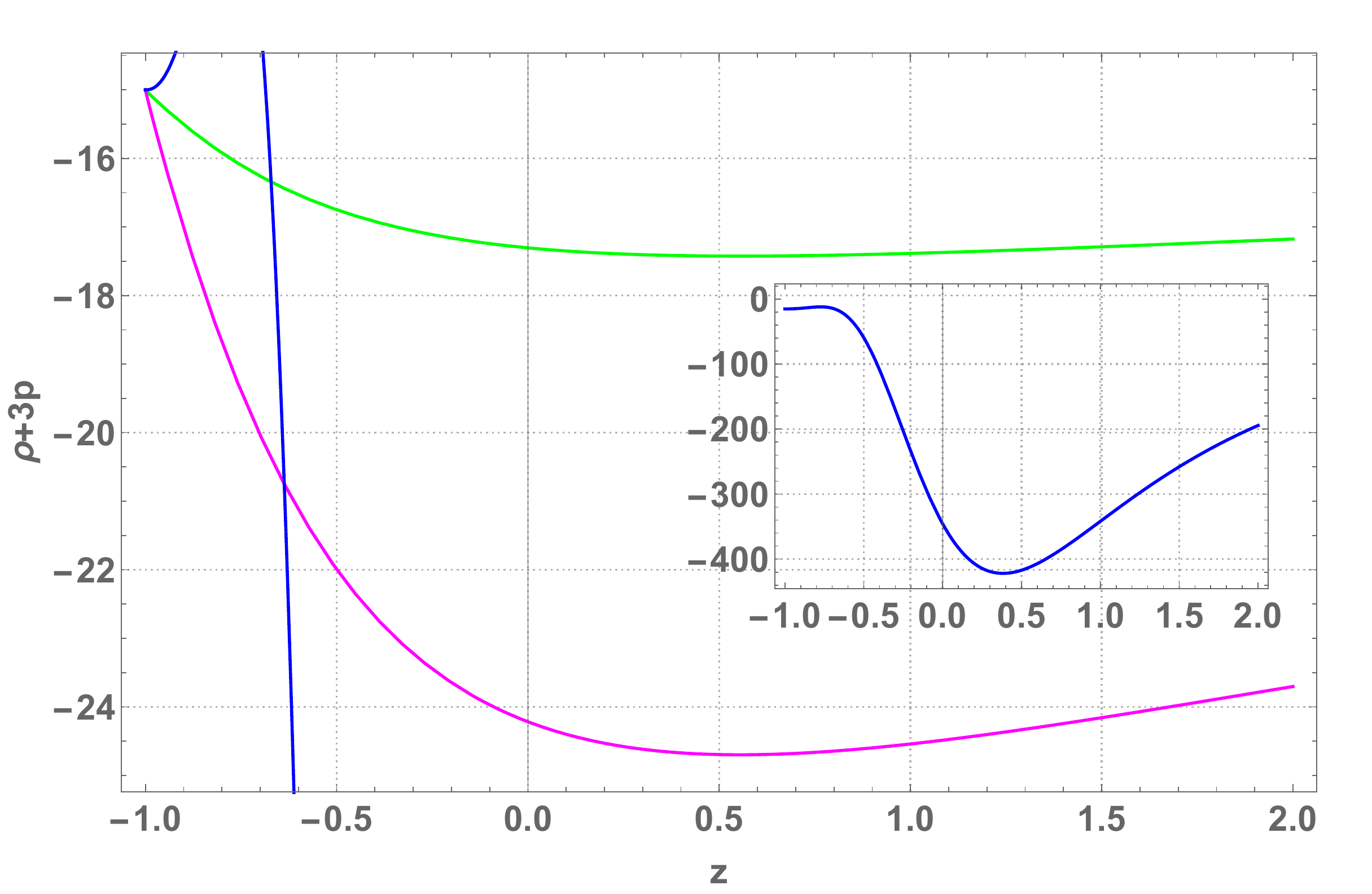}
\end{minipage}
\caption{Redshift evolution of ECs with $\alpha=-5$, $\beta=-15$,  $n=-2$, for  ($m$, $k$)=($0.480012$, $0.5$)  [green curve], ($m$, $k$)= ($0.480003$, $0.25$) [magenta curve], and ($m$, $k$)=($1.15$, $0.1$)  [blue curve]. }\label{figeccase3z}
\end{figure}

In all the above physical behavior, we experience the spacetime singularities and energy condition violations in some specific time intervals. For first two cases, the matter component and energy conditions are showing finite time singularities periodically. However, the third case has no singularities but the energy condition violations are present in some aspects. That can be understand through Raychaudhuri equations and the singularity theorems, which shows the existence of singularities in the solutions of Einstein field equations is inevitable. In this scenario, the Raychaudhuri equation can be written in term of scalar expansion $\Theta$ \cite{Wanas} as
\begin{equation}
\frac{d\Theta}{dt}= 3 \dot{H}= \frac{-3k^2 (q+1)}{m^2 \sin^2(k*t)}
\end{equation}
and the condition of non-singularity of a model is given by $\frac{d\Theta}{dt} \geqslant 0$. In herein models, the cyclic behavior of $\frac{d\Theta}{dt}$ is depicted for each set of $(m,k)$ values.    

\begin{figure}[H]
\centering

\includegraphics[width=75 mm]{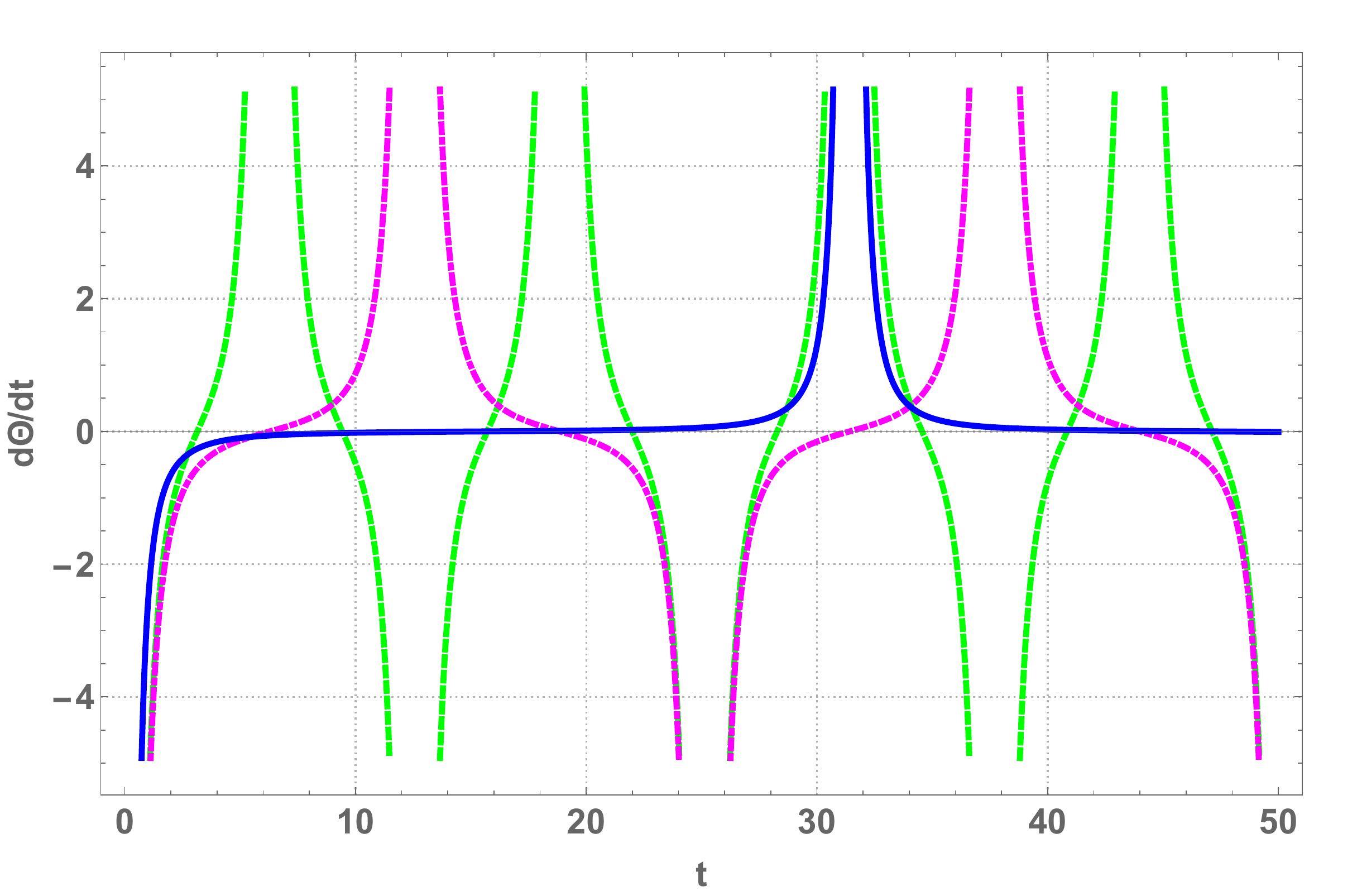}

\caption{rate of change of expansion scalar with time $t$ and $q$ for  ($m$, $k$)=($0.480012$, $0.5$)  [green curve], ($m$, $k$)= ($0.480003$, $0.25$) [magenta curve], and ($m$, $k$)=($1.15$, $0.1$)  [blue curve]. }\label{figeccase3z}
\end{figure}
 
In case of the non-singular phase, the energy conditions have to be validated as well. This can be observed in our third model that the energy conditions are violated initially due to this condition, however the model does not experience any finite time singularity within the frame work of $f(Q)$ gravity. This indicates that we can observe a non-singular cosmological model with partial violation of energy conditions within $f(Q)$ gravity formalism.

\section{Kinematical properties}\label{sec6}
In this section, we have focused on some kinematic features of our specific choice of DP. The \textit{proper distance redshift ($d(z)$)}, distance between source and observer is defined as $d(z)=a_0 r(z)$, where $r(z)$ is the radial distance of light emission and given by 
\begin{equation}\label{kpeq23}
r(z)=\int_t^{t_0} \frac{dt}{a(t)}=\int_0^z \frac{dz}{a_0H(z)}.
\end{equation}

The \textit{angular diameter distance redshift} ($d_A(z)$) represents the ratio between physical size $l$ of an object and its angular size $\theta$ ( i.e. $\frac{l}{\theta} $). The angular diameter distance $d_A$ of an object in terms of redshift $z$ is 
\begin{equation}\label{kpeq24}
d_A=\frac{d(z)}{1+z},
\end{equation}

where $d(z)$ is the proper distance redshift. Apart from these two, the another important one is known as \textit{luminosity distance redshift ($d_L$)}, which plays a vital role in observing the accelerated expansion of universe. In addition, it has important role in astronomy as it is the distance defined by the luminosity of a stellar object.  
In last few decades, modern cosmology reached a new phase to establish considerable advancements in the account of the current accelerated expanding universe. However, the observation of Type Ia Supernovae (SN Ia) is plays most significant role in this context and treated as the standard candle \cite{Perlmutter/99,Riess/98}. The use of distant SN Ia  as standard candles are considered to determine the expansion rate of universe. The Hubble constant is estimated from relative luminosity distance in terms of redshift obtained from the apparent peak of magnitude of these supernovae. Therefore the evolution of the universe is investigated by luminosity distance and this distance concepts states the explanation of expanding universe. 

Since, the distance modulus, redshift, and DP are related through Hubble parameter as a function of redshift.
It is worth to discus about the \textit{distance modulus}, which
is an observed quantity and related to luminosity distance. Here the luminosity distance ($d_L$) determining flux of the source with redshift $z$ is defined as 
\begin{equation}\label{kpeq25}
d_L=a_0 c (1+z)\int_{t}^{t_0}\frac{d\overline{t}}{a(\overline{t})}=c (1+z)\int_{0}^{z}\frac{d\overline{z}}{H(\overline{z})},
\end{equation} 
where $c$ and $a_0$ are represents the speed of light and present value of the scale factor. The distance modulus $\mu (z)$ (difference between apparent magnitude ($m$) and absolute magnitude ($M$)) in terms of $d_L$ is given as
\begin{equation}\label{kpeq26}
\mu(z)= m-M= 25+5\log_{10}\left(\frac{d_L}{Mpc}\right),
\end{equation} 
where, $d_L=c(1+z)\biggl[-\frac{2 m (1+z)^{1-m} \, _2F_1\left(1,\frac{m-1}{2 m};\frac{m-1}{2 m}+1;-(1+z)^{-2 m}\right)}{k (m-1)}\biggr]_{0}^{z}$, and $_2F_1$ is denotes the hypergeometric function.
We have obtained distance modulus $\mu(z)$  for PVDP ansatz and considered the estimated values of it from the Union 2.1 compilation supernovae datasets containing 580 points from \cite{Suzuki/12}.  In order to get the observational consistency, we have plotted  here the error bar plot for each values and found the best fit of it for $m=0.480012$ and $k=0.5$.
\begin{figure}[H]
\centering
\includegraphics[width=85 mm]{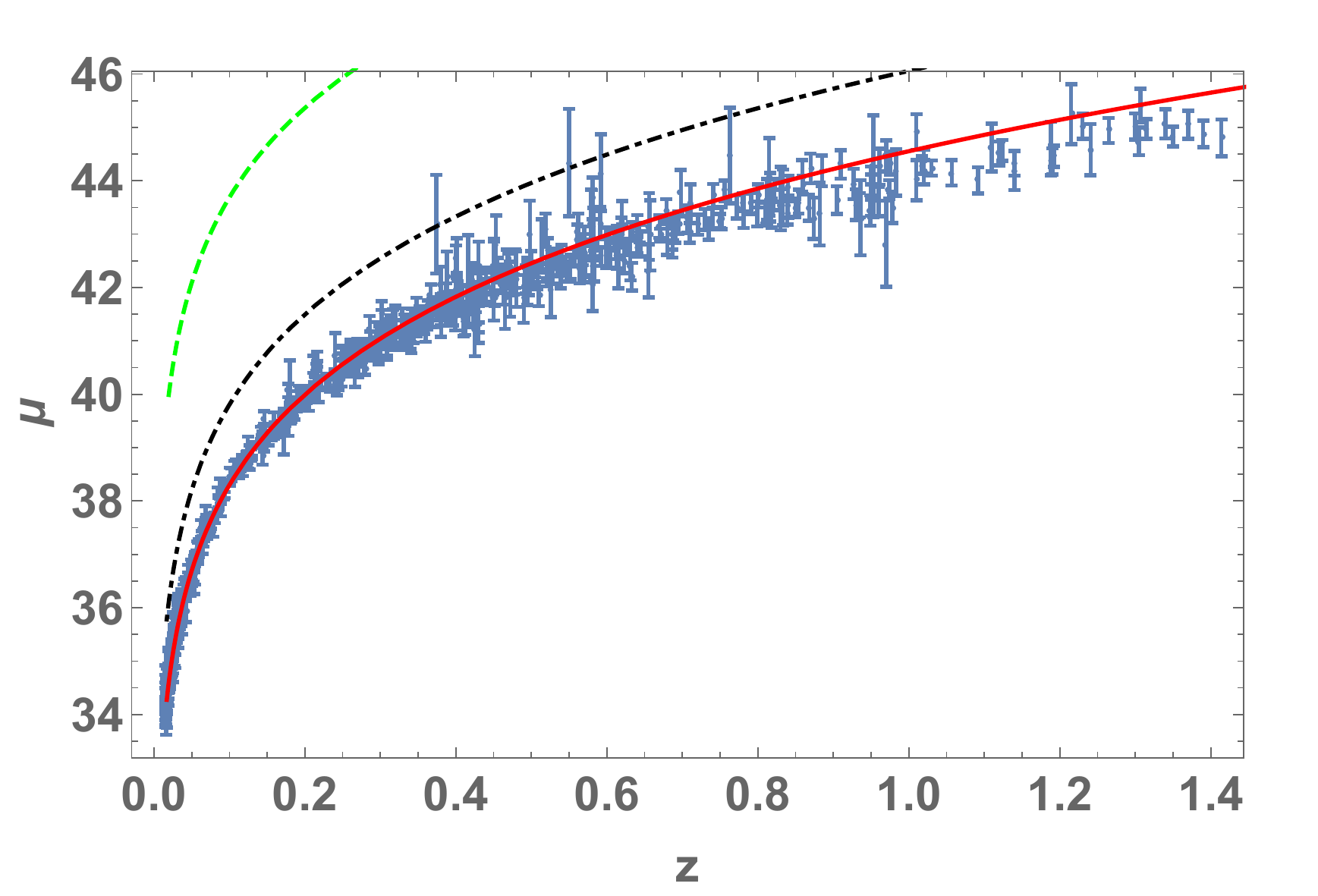}
  \caption{Error plot of distance modulus as a function of redshift for  ($m$, $k$)=($0.480012$, $0.5$)  [Red line], ($m$, $k$)= ($0.480003$, $0.25$) [Black line, DotDashed], and ($m$, $k$)=($1.15$, $0.1$)  [Green line, Dashed],  supernovae dataset [blue].}\label{SNIa}
\end{figure}

\section{Concluding Remarks}\label{sec7}

On the basis of Riemannian geometry and its extensions, three equivalent geometric descriptions of gravity can be achieved, the general theory of relativity (GR), introduced by Einstein is the first one, formulated based on a very special type of connection, the symmetric and metric-compatible Levi-Civita connection. The other two being the teleparallel and symmetric teleparallel equivalent of GR. Here, we have focused on the modified $f(Q)$ gravity or symmetric teleparallelism, where $Q$ is the non-metricity scalar. 
In this work, we have investigated the periodic cosmic evolution of $f(Q)$ gravity models. More accurately, we have concentrated on the power law form i.e. $f(Q)=\alpha Q^{n+1}+\beta$ and performed three cases for each different choices of $n=0, 1,$ and $-2$. The system of field equations have been solved with the employment of a well-tested geometric parameter called PVDP. We have used Type Ia Supernovae observational datasets to check the validity of PVDP considered in our model. We have found the best fit of it with observational data as shown in Figure \ref{SNIa}. In case of linear ($n=0$) and quadratic ($n=1$) cases, the big rip singularity have occurred at finite time and it repeats periodically. In third case ($n=-2$), the cyclic evolution has been evoked without any finite time singularity. Using the current observational values of Hubble parameter and DP and their relationship, we have performed all physical analysis for various choices of parameter values. One can observe from Figures \ref{figcase1z}, \ref{figcase2z} and \ref{figcase3z} that the energy density is positive throughout the universe whereas the pressure is always negative.  In all the three considered cases our model coincide with $\Lambda$CDM as the EoS parameter $\omega=-1$ at $z=0$. All the ECs and dynamical properties have been described with the given choices of the parameter values.
In general the modification of GR evolution usually occurs at lower curvature regime and higher curvature regime. The lower curvature regime leads to the late-time dark energy dominated universe, while the higher curvature regime is applicable for the early universe. In that sense, the EoS parameters obtained in this work have eventually showed the evolutionary phase of universe. For first two cases with $n=0$ and $n=1$, the EoS parameter value varies in between $-2.5<\omega< 0.5$, which describes the phase transition universe from radiation to matter and then dark energy era. However, in third case with $n=-2$ it lies between $-1.5<\omega<-0.5$. It defines the phase of universe completely dominated by dark energy. In addition we have found here the ECs are consistent with the periodic evolution of model EoS parameter behavior from non-phantom to phantom era in linear and quadratic case and phantom to non-phantom era in third case. 
This can be observed herewith that, the violation of ECs are mostly occurred in phantom phase. Also, the violation of ECs are leading to the existence of exotic matter fluid. That can be found in several literature \cite{i0,i1,i2,i3} and references therein, one of the best example is the traversable wormhole construction, in which the exotic matter fluid at throat of a traversable wormhole allows time travel through it. It has been observed that wormhole constructed in the framework of modified gravity theories has phantom fluid filled at its throat, which behaves as exotic to violate the ECs and allows the wormhole to be traversable. In such scenario, we can consider the phantom era of the dark energy dominated universe behaves exotic type or it has some exotic matter in it, which allows the ECs violation in certain phases. 
Henceforth, the present work contributes to the notion that cyclic behavior of late time universe can be achieved in $f(Q)$ gravity formalism with or without singularity. Also, it can be more applicable to understand the complete matter distribution of universe in $f(Q)$ gravity framework.

\section*{Acknowledgments}
PS thanks the University of Kwa-Zulu Natal for a fellowship and its continued support. 
A.D. and L.T.H. are supported in part by the FRGS research grant (Grant No. FRGS/1/2021/STG06/UTAR/02/1). We are very much grateful to the honorable referee and to the editor for the illuminating suggestions that have significantly improved our work in terms of research quality, and presentation.

\end{document}